\documentclass[iop]{emulateapj}
\usepackage[flushleft]{threeparttable}
\usepackage{amsmath}

\usepackage{float}

\usepackage{mathrsfs}

\begin{document}

\title{Characterization of Low Mass K2 Planet Hosts using Near-Infrared Spectroscopy}

\author{Romy~Rodr\'iguez~Mart\'inez\altaffilmark{1}}
\author{Sarah~Ballard\altaffilmark{2,3}}

\altaffiltext{1}{Department of Astronomy, The Ohio State University, 140 West, 18th Avenue, Columbus, OH 43210, USA; rodriguezmartinez.2@osu.edu}
\altaffiltext{2}{MIT Kavli Institute for Astrophysics \& Space Research, Cambridge, MA 02139, USA}
\altaffiltext{3}{MIT Torres Fellow}

\author{Andrew~Mayo\altaffilmark{4,5}}

\altaffiltext{4}{DTU Space, National Space Institute, Technical University of Denmark, Elektrovej 327, DK-2800 Lyngby, Denmark}
\altaffiltext{5}{Centre for Star and Planet Formation, Natural History Museum of Denmark \& Niels Bohr Institute, University of Copenhagen, \O ster Voldgade 5-7, DK-1350 Copenhagen K., Denmark}
\altaffiltext{$\ddagger$}{National Science Foundation Graduate Research Fellow}
\altaffiltext{$\star$}{Fulbright Fellow}

\author{Andrew~Vanderburg\altaffilmark{6,7}}

\altaffiltext{6}{Department of Astronomy, The University of Texas at Austin, Austin, TX 78712, USA}
\altaffiltext{7}{NASA Sagan Fellow}

\author{Benjamin T.~Montet\altaffilmark{8,9}}

\altaffiltext{8}{Department of Astronomy and Astrophysics, University of Chicago, 5640, S. Ellis Ave, Chicago, IL, 60637, USA}
\altaffiltext{9}{NASA Sagan Fellow}

\author{Jessie L.~Christiansen\altaffilmark{10}}

\altaffiltext{10}{NASA Exoplanet Science Institute, California Institute of Technology, M/S 100-22, 770 S. Wilson Ave, Pasadena, CA, USA}


\begin{abstract}
We present moderate resolution near-infrared spectra in $H, J$ and $K$ band of M dwarf hosts to candidate transiting exoplanets discovered by NASA's K2 mission. We employ known empirical relationships between spectral features and physical stellar properties to measure the effective temperature, radius, metallicity, and luminosity of our sample. Out of an initial sample of 56 late-type stars in K2, we identify 35 objects as M dwarfs. For that sub-sample, we derive temperatures ranging from 2,870 to 4,187 K, radii of $0.09-0.83$ $R_{\odot}$, luminosities of $-2.67<log L/L_{\odot}<-0.67$ and [Fe/H] metallicities between $-0.49$ and $0.83$ dex. We then employ the stellar properties derived from spectra, in tandem with the K2 lightcurves, to characterize their planets. We report 33 exoplanet candidates with orbital periods ranging from 0.19 to 21.16 days, and median radii and equilibrium temperatures of 2.3 $R_{\oplus}$ and 986 K, respectively. Using planet mass-radius relationships from the literature, we identify 7 exoplanets as potentially rocky, although we conclude that probably none reside in the habitable zone of their parent stars. 

\end{abstract}

\section{Introduction}

Since its launch in 2009, the NASA $Kepler$ spacecraft has gathered exquisite photometry of over 150,000 stars and has uncovered thousands of exoplanets in our galaxy via the transit photometry method \citep{2010Sci...327..977B,2011ApJ...728..117B,2013ApJS..204...24B,2014yCat..22100019B}. $Kepler$ continuously monitored the same part of the sky for four years, until reaction wheel failure compromised the pointing stability of the spacecraft. However, engineers soon found a way to balance the spacecraft using solar pressure and repurposed it for a new mission, K2 \citep{2014PASP..126..398H}. In this new mode of operation, K2 observes different regions along the ecliptic, targeting between 10 and 30 thousand stars for approximately 80 days. K2 is therefore particularly suited for searches of transiting exoplanets in short-period orbits.

The motivations for targeting M dwarfs for both exoplanet searches and follow-up observations are manifold. First, M dwarfs are the most common type of star, comprising nearly 70\% of all stars in the Milky Way \citep{2010AJ....139.2679B}. Second, although they were initially thought to host planets infrequently for their dearth of Jupiter-size planets, the $Kepler$ and K2 missions revealed that M dwarfs form smaller (potentially rocky) planets in greatest abundance \citep{2012ApJS..201...15H}. Studies have shown that for planets with periods of less than 50 days, planets between $2-4R_{\oplus}$ are twice as abundant around M dwarfs than around sunlike stars \citep{2012ApJS..201...15H,2015ApJ...814..130M}. This fact, combined with the ubiquity of M dwarfs, establish them as the majority of hosts to small planets in the Milky Way. \citet{2015ApJ...807...45D} found that the mean number of small planets $(0.5-4R_{\oplus})$ per late K dwarf or early M dwarf is $2.5\pm0.2$ for orbital periods shorter than 200 days, comparable to the $2.0 \pm 0.45$ determined by \citet{2014DPS....4630109M}. Additionally, the smaller radii and masses of M dwarfs translate to larger transit depths, larger radial velocity semi-amplitudes, and larger transmission spectroscopy signals for exoplanet study (for a detailed summary of the advantages and complications of M dwarfs as planet host stars, see \cite{2016PhR...663....1S}).
The recent discoveries of small, temperate exoplanets circling M dwarfs, such as Proxima b \citep{2016Natur.536..437A}, the TRAPPIST-1 system \citep{2017Natur.542..456G} and LHS1140 b \citep{2017Natur.544..333D} have further demonstrated the feasibility of targeting cool, small stars in the search for potentially life-bearing worlds.

Despite these facts, there is a relative paucity of detected planets orbiting M dwarfs. They number several hundred, as compared to the several thousand of their FGK counterparts
\footnote[1]{https://exoplanetarchive.ipac.caltech.edu/}. They are challenging to characterize from spectra (\citet{2011ESS.....2.1903T}, with summary in \citet{2016PhR...663....1S}) and also comprised a small fraction in the $Kepler$ Input Catalogue \citep{2011AJ....142..112B}. However, recent studies have made critical inroads linking spectral features to physical properties of M dwarfs \citep{2012ApJ...757..112B, 2012ApJ...748...93R, 2012ApJ...747L..38T, 2012ApJ...753...90M, 2013ApJ...779..188M, 2015ApJ...800...85N}. Moreover, stars cooler than 4000 K make up 25\% of the TESS Input Catalog \citep{2015ApJ...809...77S,2018AJ....155..180M}, as opposed to 5\% of the $Kepler$ Input Catalog \citep{2011AJ....142..112B}.

Precise stellar characterization is ultimately crucial to understand the planet sample. The characteristics of these new worlds are so closely tied to the physical properties of their host stars, that we must understand the stars first if we aim to understand the planets in detail. Eking out the mass-radius relationship of exoplanets, for example, relies on large spectroscopic or asteroseismic surveys to characterize the host stars to better than $10\%$ \citep{2013AAS...22140705W,2013giec.conf30103H,2014AJ....148...78D,2016ApJ...825...19W,2017AJ....154..109F,2018MNRAS.479.4786V}. Furthermore, the deluge of exoplanet discoveries and our limited resources make it impossible to follow-up every single planet candidate.  Reliably identifying the most promising candidates for follow-up characterization, demands that we know the characteristics of the candidates.

For K2, in contrast to $Kepler$, the target-selection has been proposal-driven. Likewise, stellar characterization of large samples of K2 planet host stars has been an ongoing community effort \citep{2016ApJS..224....2H,2017ApJ...836..167D}. To contribute to this endeavor, we present in this study the stellar characterization of 35 candidate exoplanet host stars from K2 with near-infrared spectra.

We infer the temperatures, radii, luminosities, mass and metallicities of the stellar sample using empirial relationships. We subsequently estimate the radii and equilibrium temperatures of the planet candidates. The paper is organized as follows: in Section 2, we describe the observation techniques and the reduction pipeline of our spectra. In Section 3, we present an analysis of the data, the derivation of the equivalent widths (EWs) of different metals and a comparison of our derived equivalent widths of the aluminum feature at 1.67 microns to those previously published. In Section 4, we summarize the results of our analysis for the cool dwarf sample. In Sections 5 and 6, we explain the derivation of the planet parameters and discuss the potential habitability of the planets. We identify two systems suitable for follow-up Doppler and atmospheric characterization, and we highlight several false positives from the K2 photometry pipeline. In Section 7, we conclude and summarize our findings and recommendations for follow-up observations. 

\section{Observations} \label{sec:observations}

We gathered our spectra with the near-infrared TripleSpec spectrograph \citep{2004SPIE.5492.1295W} at the Palomar 5m telescope. The TripleSpec instrument has a $1"\times30"$ slit, a moderate resolution $R = 2500-2700$ and a wavelength coverage in the NIR of 1.0 to $2.4  \mu m$. We selected our sample of 56 late-type stars identified as potential planet hosts from the first two years of the K2 mission.
For each science target, we gathered observations in an ABBA nod sequence, with exposure times between 5 and 300 seconds. We also gathered spectra of telluric standards at hourly intervals. We used the bright quartz lamp to gather flat fields at the beginning and end of each night, in addition to collecting dark frames.

To reduce our spectra, we used Spextool, a publicly available IDL-based package for spectral reduction \citep{2004PASP..116..362C}. For correction of telluric lines, we used xtellcor \citep{2003PASP..115..389V}. When choosing a spectral line in the A0 star that is unaffected by atmospheric absorption, we selected the Paschen $\delta$ line at $1.005\mu m$, following the TripleSpec manual suggestions.

The observing conditions varied slightly 
throughout the night from mostly clear skies and light wind early in the night, to considerably cloudy (cirrus clouds) with heavy humidity and fog. Weather variability affected the observations to the point where some of the spectra were too low signal-to-noise to yield reliable estimates of the stellar parameters, and are therefore not included here. Additionally, we removed all the objects with exposures of 300s or more, as the sky was too variable over that duration, making the sky subtraction challenging.

We also rejected all the reduced stars with derived temperatures below 2,800 K or above 4,800 K since those are outside the range for which the empirical relationships for cool dwarfs used here are valid.

\subsection{Contamination by Red Giants} \label{sec:contamination}

One of the challenges in characterizing M dwarfs from $Kepler$ and K2 samples is contamination by red giants. 
Because the target selection from $Kepler$ and K2 is based on color, many samples of putative low-mass red stars are polluted by red giants or hotter stars reddened by interstellar extinction. Mann et al. 2012 found that, from a sample of 382 supposed M dwarfs from $Kepler$, the majority were in fact giants. This suggests that the identification of M dwarfs from photometry alone in large samples is unreliable or impossible without spectroscopic follow-up.
We therefore began our analysis of the sample by visually inspecting all the spectra to eliminate the red giants or stars with other classifications. We compared features that appear sharply distinct in red giants and red dwarfs. Some of the most prominent relative differences in the $J$, $H$ and $K$ bandpasses in dwarfs and giants are in the lines Mg $(1.50\mu m)$, Mg $(1.71\mu m)$ and Na at $2.2 \mu m$. The results of this analysis show that, from a total of 56 red stars, only 35 are M dwarfs, while the rest were either visually classified as red giants, or their spectra were consistent with late K or hotter dwarfs. Some of those stars include EPIC 210769880, classified as a K2 giant in \citet{2017ApJ...836..167D}, EPIC 211762841, a K7 dwarf as classified in the same work, and EPIC 211822797, another K7 dwarf. Finally, EPIC 211694226 is classified in \citet{2017ApJ...836..167D} as an M3 dwarf with a nearby companion which may or may not be physically associated. When reducing the spectrum of this star, the light from the companion contaminated it and we could not properly reduce it so we exclude it from the characterization.
 
\section{Analysis}
\label{sec:analysis}

To extract the physical properties of M dwarfs from their spectra, we cannot directly compare whole synthetic spectra to observed spectra as for FGK dwarfs (see \citet{2016PhR...663....1S} review for detailed summary.) Current best practices involve empirical relationships between physical properties and spectral features, painstakingly acquired with benchmark binary systems or interferometric measurements. With the exception of \citet{2013AJ....145...52M}, in which the authors identified large sections in which synthetic spectra do reliably replicated observed spectra, most published relationships employ spectral indices and equivalent widths of absorption features across the optical and near infrared \citep{2012ApJ...748...93R,2012ApJ...747L..38T,2012ApJ...753...90M,2015ApJ...800...85N}. We note that the equivalent width of a spectral feature is defined as

\begin{equation}
EW_{\lambda} = \int_{\lambda_{1}}^{\lambda_{2}} [1 - \frac{F(\lambda)}{F_{c}(\lambda)}]\ d\lambda 
\end{equation}

where $F(\lambda)$ is the flux of the absorption line integrated between $\lambda_{1}$ and $\lambda_{2}$ and $F_{c}(\lambda)$ is the continuum flux. 

\begin{figure}
\centering 
\includegraphics[width=\columnwidth]{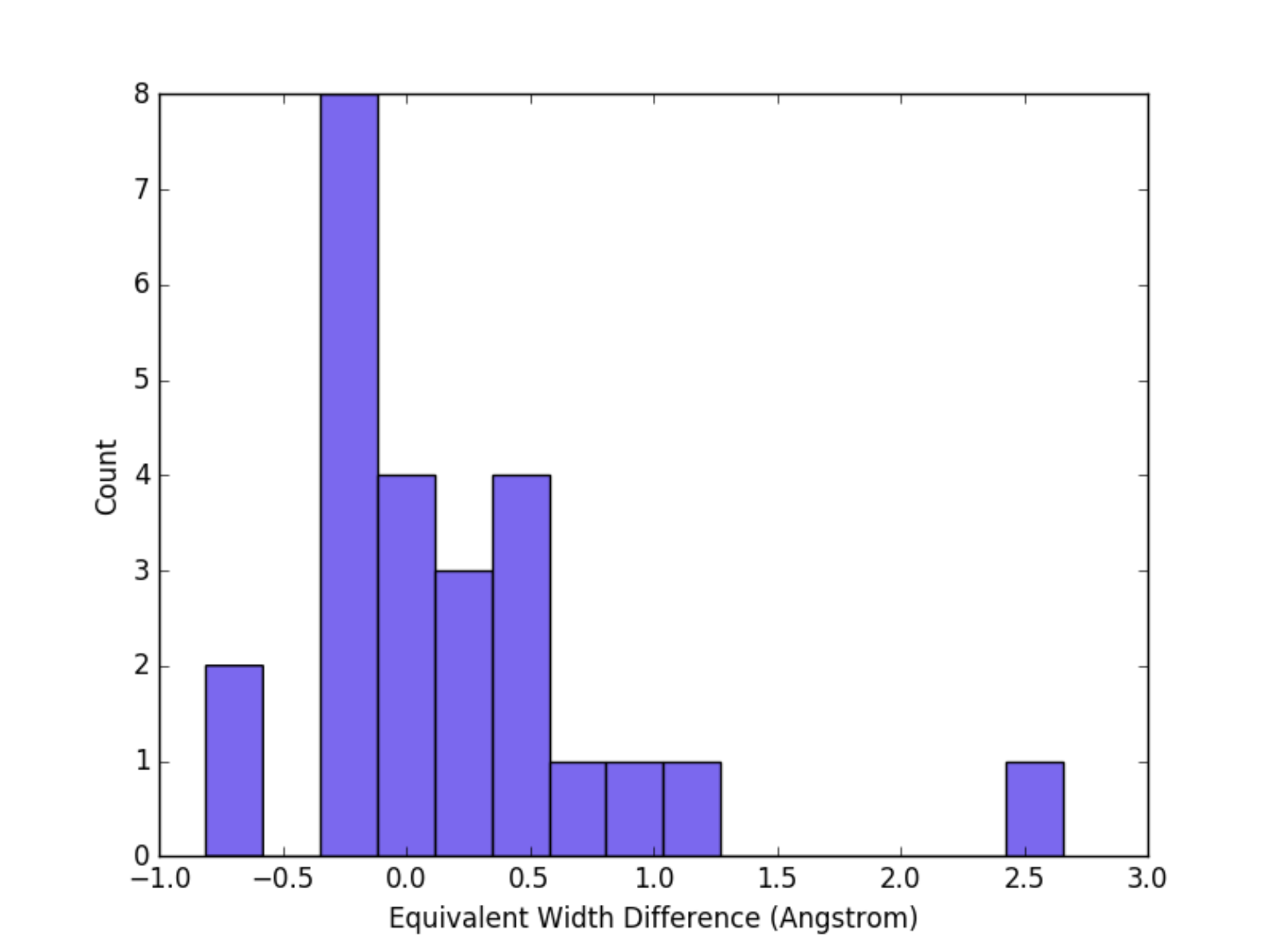}
\caption{Difference between two distributions: the equivalent widths of the line $Al-a (1.67 \mu m)$ (which is sensitive to both temperature and radius) of the cool dwarfs in this paper and the equivalent widths of that same spectral line for those same stars previously characterized in \citet{2017ApJ...836..167D}.}
\label{DR}
\end{figure} 

We elect to employ here the relationships of \citet{2015ApJ...800...85N} to derive stellar radius from the spectrum, and \citet{2013ApJ...779..188M} for deriving metallicity. We selected the metrics with smallest intrinsic scatter, which are $H$ band from \citet{2015ApJ...800...85N} for stellar radius and effective temperature, and $K$ band from \citet{2013ApJ...779..188M} for metallicity. The relationships from \citet{2015ApJ...800...85N} are generally applicable to spectral types between  mid-K and mid-M, with radii of $0.18 <R/R_{\odot} < 0.8$, temperatures of $3,200 K <T_{\rm{eff}}<4,800 K$, and log luminosities of $-2.5<log(L/L_{\odot})< -0.5$. 

In that work, they determined that stellar effective temperature correlated most strongly with the equivalent widths of the aluminum doublet at $1.67 \mu m$ and magnesium absorption at $1.50 \mu m$, with intrinsic scatter of 73 K. The same aluminum doublet, in addition to magnesium absorption at $1.57 \mu m$, traced stellar radius, with intrinsic scatter of $0.027 R_{\odot}$. We also measure stellar luminosity, which \citet{2015ApJ...800...85N} found to be correlated with the equivalent widths of magnesium features at 1.50 and $1.71 \mu m$, with intrinsic scatter of 0.049 in $log (L/L_{\odot})$.

\citet{2013AJ....145...52M} also used the EWs of absorption features to determine the [Fe/H] and [M/H] metallicity of cool dwarfs. They found that the lines most sensitive to [Fe/H] metallicity are features in $K$ band, including Na at $2.2 \mu m$. Stellar metallicity is also related to the spectral index H$_{2}$O-K2 introduced by \citet{2012ApJ...748...93R}, which measures the deformation of the spectrum in $K$ band due to water absorption, and is temperature sensitive. Because the index saturates at temperatures close to 4,000 K, the metallicity relations from \citet{2013AJ....145...52M} are not reliable for hotter stars.

We went on to calculate stellar mass using another empirical relation reported in \citet{2013ApJ...779..188M}. This relation is a third degree polynomial with effective temperature. We use the newly derived effective temperature from the \citet{2015ApJ...800...85N} metric to then obtain the stellar mass.

\citet{2013ApJ...779..188M} caution that these relationships for single stars give increasing errors for stars with $T_{\rm eff} <$ 3,300 K. Because there are several stars below this 
temperature in our sample, some of the stellar masses yielded negative values as they had very low effective temperatures. For the stars for which the \citet{2013ApJ...779..188M} relationships yielded negative masses, we employed the mass-radius empirical relationships for single stars (equation 10) from \citet{2012ApJ...757..112B}.

We computed EWs by first selecting a region to the left and to the right of the spectral feature of interest (both the line centers, and the regions on either side used to normalize the flux, are listed in \citet{2015ApJ...800...85N}). We then numerically integrated the flux within the absorption feature, and we define  the EWs in Angstrom. The derived values for the EWs and masses are shown in Table 2. We tested the numerical approach of measuring equivalent widths by fitting Gaussian curves to the absorption lines with Gaussian profiles. The values obtained from both methods are in good agreement, though we elected to use the numerical method to measure EWs of lines with non-Gaussian profiles (such as the aluminum doublet at $1.67 \mu m$). Figure 1 depicts the difference between our derived EW values of $Al-a (1.67 \mu m)$ for the set of 24 stars in our sample that overlap with \citet{2017ApJ...836..167D}. We are consistent with that work to within 0.5{\AA} with 68\% confidence.

We employed a bootstrap technique to measure the uncertainties on our equivalent widths and the corresponding stellar parameters. We generated 100 synthetic spectra from each individual spectrum with the same noise properties as the real spectra, then calculated all the parameters of interest in each of them. We then took the mean as the true measured value and the standard deviation as the error associated to each stellar property.

\section{Results} \label{sec:results}

\subsection{Spectral Type} \label{sec:spectraltype}

\begin{figure}
\centering
\includegraphics[width=0.8\columnwidth]{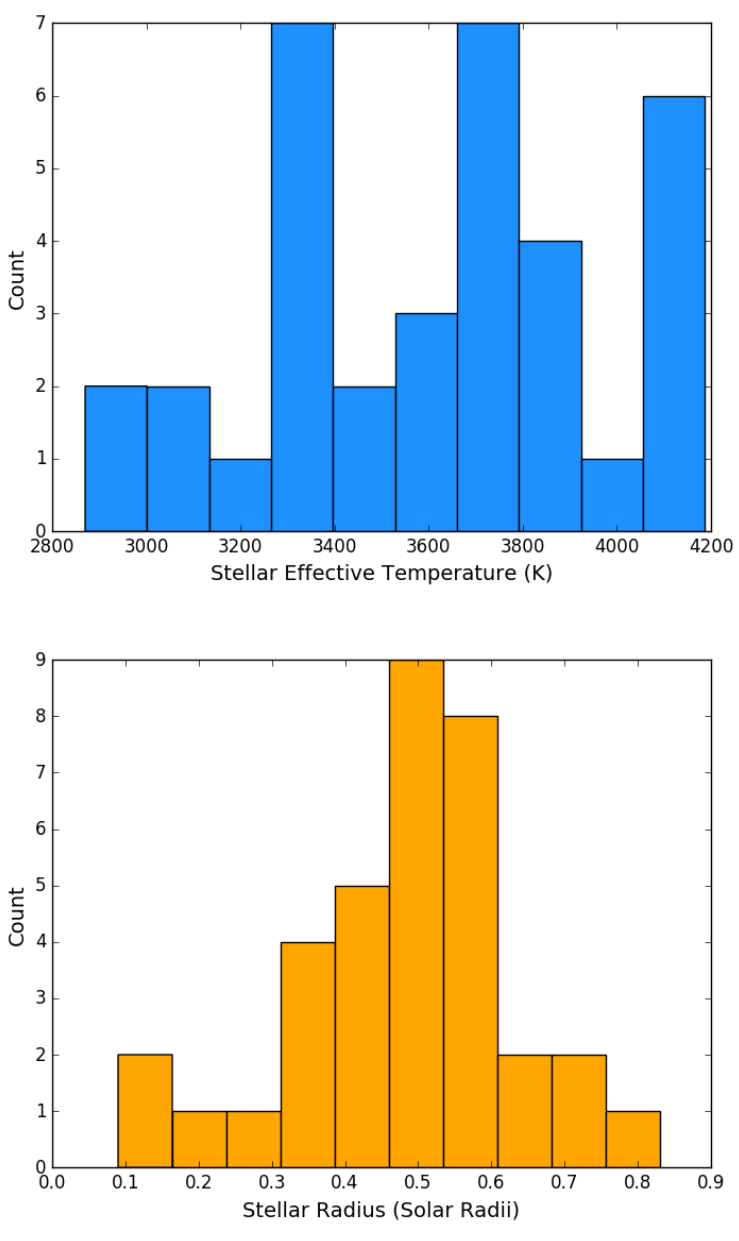}
\caption{Temperature (top) and radius (bottom) distribution of our stellar sample. The mean temperature is 3,681 K and the median uncertainty is around 100 K. The radius distribution peaks sharply at 0.48 $R_{\odot}$.}
\label{TS}
\vspace{5mm}
\end{figure}

By applying the empirical calibrations from \citet{2015ApJ...800...85N} for temperature, and by the measuring the EWs of $Al-a (1.67 \mu m)$, $Al-b (1.67 \mu m)$ and  $Mg (1.50 \mu m)$ lines, we estimated the stellar effective temperature for all the stars for which we had reduced spectra. The temperatures for our sample range from 2,870 K to 4,187 K, which brackets spectral types between M5 and K7 \citep{2005nlds.book.....R,2012ApJ...757..112B}. For stars with negative EWs or stars cooler than about 2,800 K or hotter than 4,800 K, we discarded them from the characterization for being outside the range for which the \citet{2015ApJ...800...85N} relationships are valid. Figure 2 shows a histogram of the distribution of temperatures of our cool dwarfs. The mean value of the distribution is 3,620 K. The median uncertainty in the stellar effective temperature is about 100 K. The error bars in the temperatures of many of the stars is large in part because of the disappearance of the Mg and Al lines in the coolest dwarfs, which causes the temperatures to be overestimated \citep{2015ApJ...800...85N,2017ApJ...836..167D}. 

\subsection{Stellar Radius} \label{sec:stellar radius} 

We estimated the radii of all the stars in our sample applying the \citet{2015ApJ...800...85N} relationships by calculating the EWs of the sensitive lines to radius. We obtained a range of radii between 0.09 $R_{\odot}$ and 0.83 $R_{\odot}$. The mean of the distribution is 0.48 $R_{\odot}$. The mean error on our sample is 0.05 $R_{\odot}$. Figure~\ref{TS} shows the radii distribution. As with the temperatures, the range of measured radii is consistent with an M dwarf classification. Some of the equivalent widths sensitive to stellar radius were negative, yielding radii too low for the \citet{2015ApJ...800...85N} relationships. This happens likely because of a flattening of the continuum as the temperature decreases for very late M dwarfs.  

\subsection{Luminosity}

The $log L/L_{\odot}$ luminosities from our stellar sample were all estimated using the \citet{2015ApJ...800...85N} relations, and yielded a range of luminosities between $-2.67 < log L/L_{\odot}<-0.67$, or between 0.002  $L_{\odot}$ and 0.213 $L_{\odot}$. Figure 4 shows the range of values for the luminosities. The average log luminosity is -1.59, and the mean error is 0.13.

\begin{figure}
\centering 
\includegraphics[width=\columnwidth]{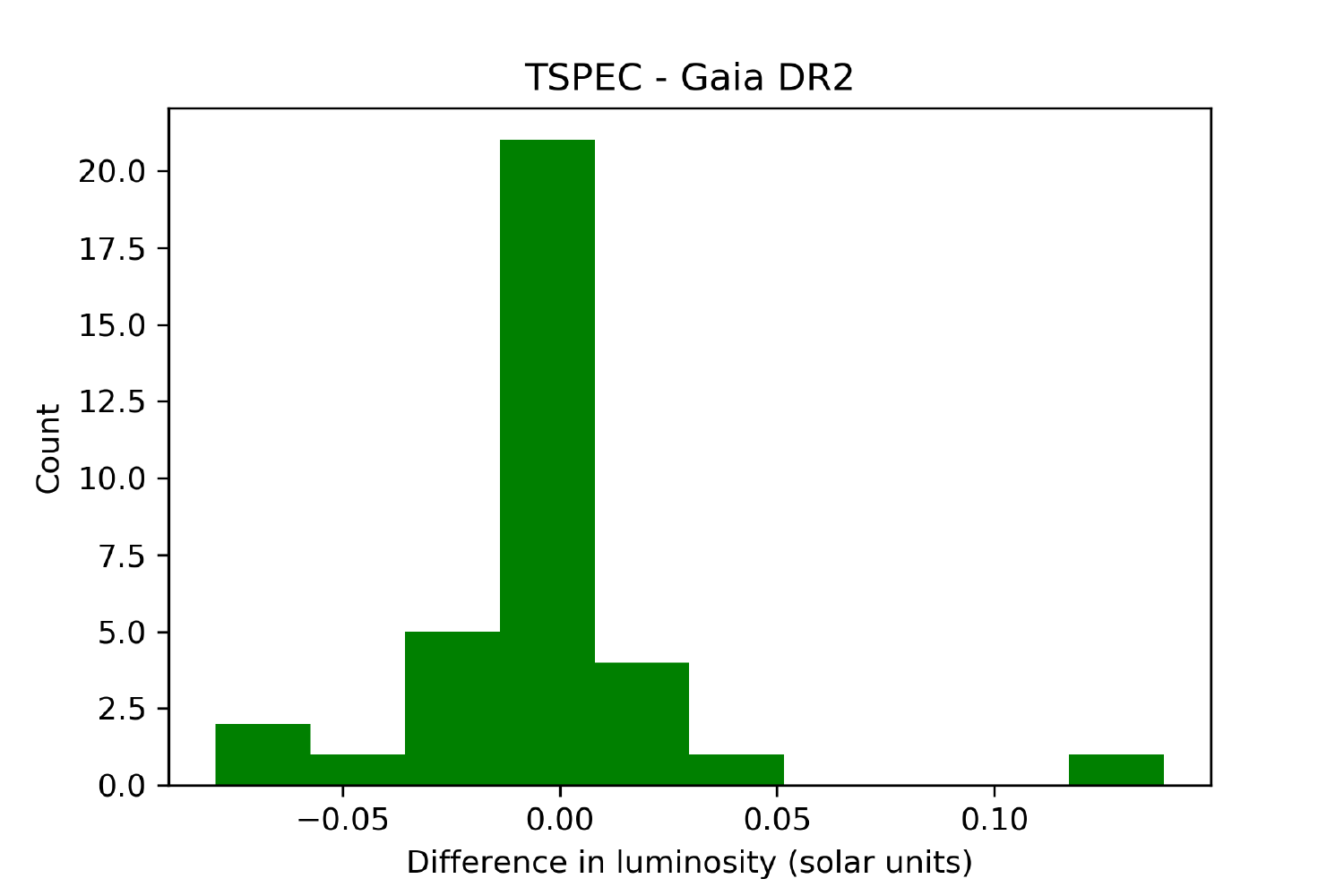}
\caption{Difference between the distribution of the luminosity values (in solar units) of our stellar sample and those predicted by Gaia Data Release 2. The underluminous outlier in the distribution is the early M dwarf EPIC 201635569.}
\label{gaia}
\end{figure} 

\subsection{Metallicity}

The abundance of metals in a star can predict whether it will form planets, and the types of planets it will form \citep{2004A&A...415.1153S,2005ApJ...622.1102F,2009ApJ...699..933J}. Studies indicate that a higher metallicity is directly related to more matter in the protoplanetary disk, and therefore there is a higher chance of making more gaseous and rocky planets. Authors in recent years have already hinted at correlations between metallicity and eccentricity \citep{2013ApJ...767L..24D}, and mutual inclinations and planet multiplicity \citep{2016ApJ...816...66B}. \citet{2018AJ....155..127H} found that from a sample of 16 planets orbiting M dwarfs, those larger than 3 $R_{\oplus}$ are only found orbiting the most metal-rich hosts.

We derived [Fe/H] metallicities in our sample from the \citet{2013AJ....145...52M} relationships by calculating the EWs of the features sensitive to metallicity in $K$ band and by computing the H$_{2}$O-K2 index in the specified range in that same paper. The metallicities of our sample of M dwarfs ranged from -0.49 to 0.83 dex. The mean uncertainty in our values is 0.1 dex, as can be observed in Figure 4. The range of host star metallicities in our sample is similar to that of surveys of hundreds of nearby M dwarfs from both \citet{2012ApJ...748...93R} and \citet{2014AJ....147...20N}. 
All the derived physical parameters and uncertainties for our cool dwarfs are listed in Tables 1 and 2. 

\begin{figure}
\centering
  \includegraphics[width=\columnwidth]{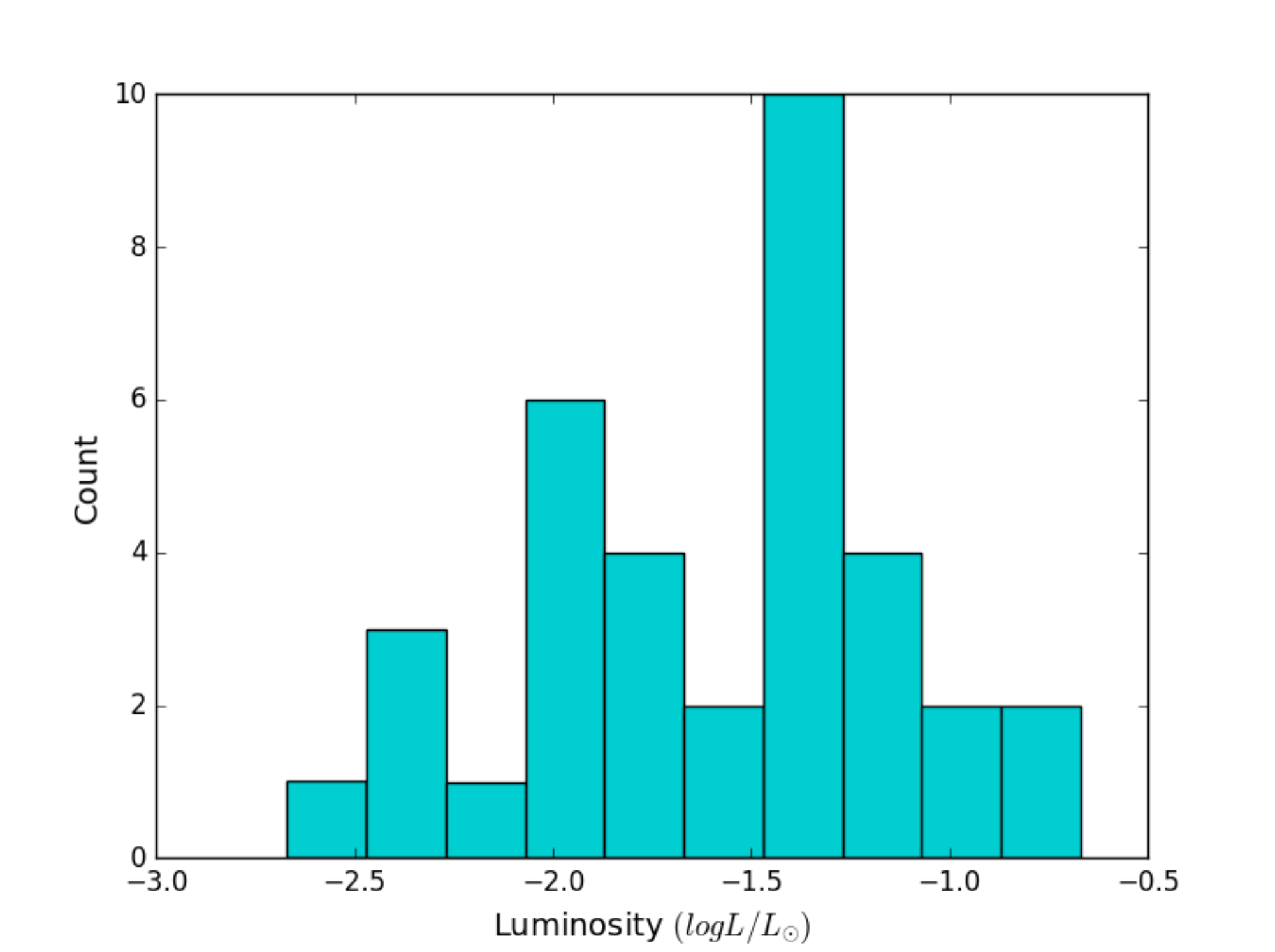}
  \includegraphics[width=\columnwidth]{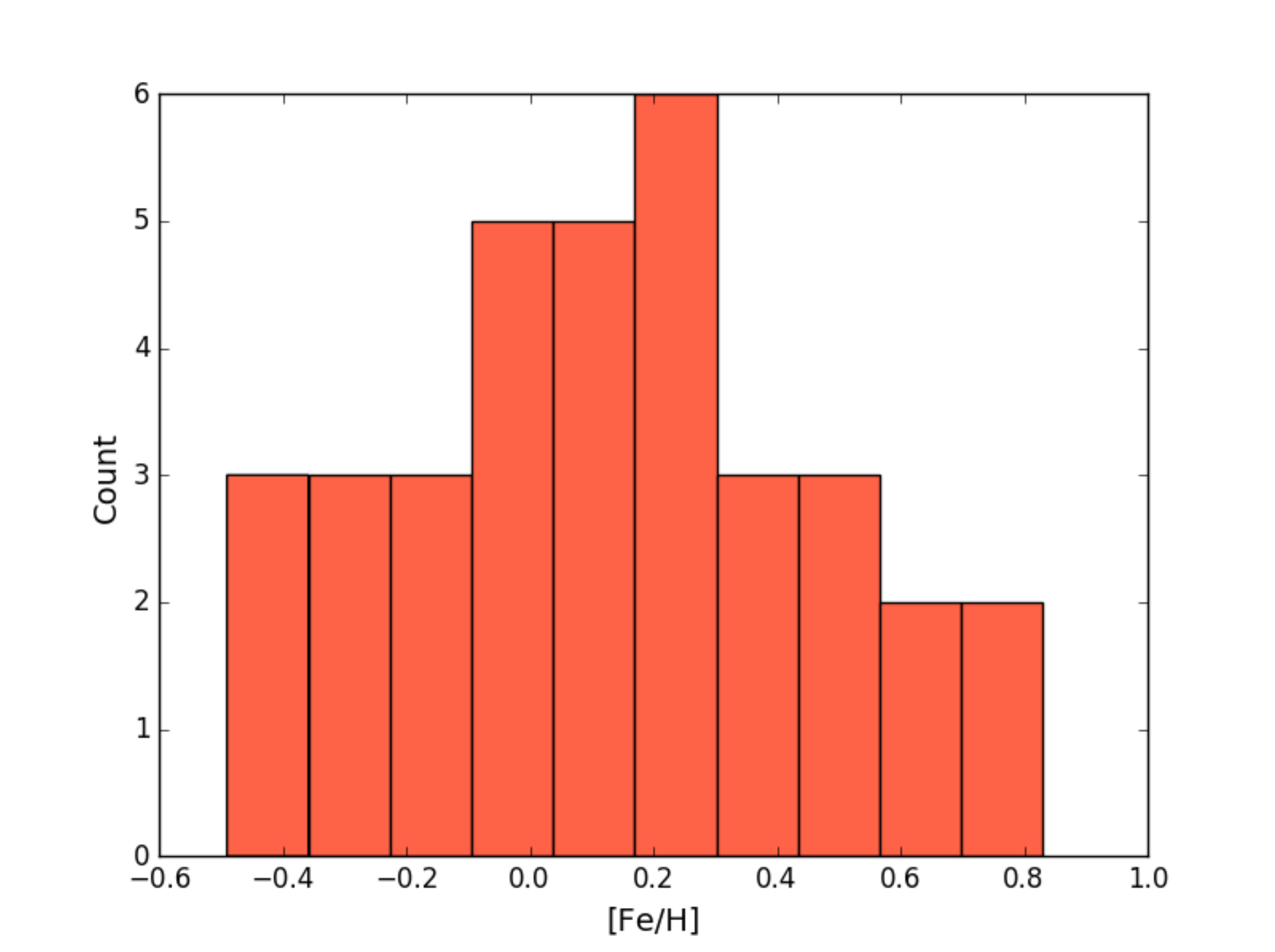}
\caption{Luminosity (top) and [Fe/H] metallicity (bottom) of our sample. The median log luminosity is -1.57 while the mean metallicity is 0.13 dex.}
\label{LMet}
\end{figure}

\subsection{Comparison of Stellar Parameters with Gaia DR2} \label{sec:gaia}

We complemented the observations of our sources with the recent second data release of the Gaia mission \citep{2016A&A...595A...1G,2018arXiv180409365G}. We corroborated the accuracy of our derived stellar parameters by comparing our luminosities from the \citet{2015ApJ...800...85N} empirical relationships to those predicted by Gaia DR2. First, we computed the absolute magnitude of each star using the exquisitely precise G-band magnitudes and distances inferred from the parallaxes delivered by Gaia. We then converted the absolute G magnitudes to bolometric absolute magnitudes by applying the bolometric corrections for cool dwarfs in \citet{2010A&A...523A..48J}. Finally, we converted the bolometric absolute magnitudes to bolometric luminosities and compared them with the luminosities estimated in this work. We found no evidence of offset between the distributions, and conclude that our results are in good agreement with Gaia, as can be observed in Figure~\ref{gaia}.  

\section{The Planet Sample} \label{sec:sample}

\subsection{Transit Lightcurve Reduction and Analysis}

In addition to the stars, we characterized the associated planet candidates using the derived stellar properties presented here. Specifically, we applied the inferred stellar radii, luminosities and masses to estimate planet radius, equilibrium temperature and semi-major axis, respectively. We briefly describe the process of reduction and analysis of the K2 lightcurves of our stellar sample and how we estimated the planet properties.

Before fitting and analyzing the lightcurves for the transit and planetary parameters, the K2 photometry must be corrected for motion-induced systematics that reduce its photometric precision and introduce noise and artificial variability in the data. We compensate for such instrumental systematics using the reduction strategy outlined in \citet{2014PASP..126..948V} and later updated in \citet{2016ApJ...829L...9V}. The first step of this process consists in creating 20 aperture masks of varying sizes and shapes to perform aperture photometry on the K2 targets and produce 20 raw lightcurves of each. The systematic noise due to the loss of balance and the regular repositioning of the spacecraft is then mitigated essentially by estimating the path of the targets along the CCD and identifying and removing a correlation between $Kepler$ motion and the apparent measured flux. Additionally, we identify and remove the data taken during thruster fires and also eliminate low-frequency ($>0.75$ days) variations. After implementing these steps on all 20 lightcurves, we pick the aperture mask that generates the lightcurve with the greatest photometric precision and quality.

Following the correction of systematics and removal of low-frequency variations in the lightcurves, we proceeded to reproduce the original lightcurves to assess the final transit and orbital parameters for our candidates. We estimated the systematic errors in our final lightcurve by fitting for them simultaneously with the transit parameters and low variability in the lightcurve. The systematics were modeled as a spline in arclength, or $Kepler$ position in its roll, and the low-frequency variability was modeled as a basis spline in time, with break points every 0.75 days.
The lightcurve transit parameters were modeled with the transit model from \citet{2002ApJ...580L.171M}, while the fits to the transits were done using a Levenberg-Marquardt optimization \citep{2009ASPC..411..251M}. We further analyze the lightcurves using the BATMAN Python package from \citet{2015PASP..127.1161K} to calculate the model transit lightcurves and thus estimate the final transit parameters and uncertainties. We assumed that all the planetary candidates were non-interacting. We included five parameters for each candidate: the time of the first transit, the period, inclination, the ratio of planet to star radius ($R_{p}/R_{\star}$), and semi-major axis normalized to stellar radius ($a/R_{\star}$). Finally, the transit parameters in the model were estimated using $emcee$ \citep{2013PASP..125..306F}, a Python package which simulates lightcurves using a Markov chain Monte Carlo algorithm. For a more thorough description of this procedure, see \citet{2018AJ....155..136M}. 

\subsection{Planet Properties} \label{sec:planetproperties}

For the calculation of the equilibrium temperature, we assume zero eccentricity and an albedo of zero. The median equilibrium temperature in the sample is 986 K, while the majority of planets have radii between 2 $R_{\oplus}$ to 4 $R_{\oplus}$ rendering them “Neptune-Size” planets, consistent with the known planet size demographic of M dwarfs, most of which host sub-Neptune planets \citep{2013ApJ...767...95D,2015ApJ...807...45D,2014ApJ...791...10M}. All the derived planet parameters are listed in Table 3. Figure 5 shows the distribution of sizes into four typical exoplanet size categories. We found two interesting outliers in the sample: EPIC 211995398 and EPIC 211509553, which show deep transits of $R_{p}/R_{}\star = 0.15$ and $R_{p}/R_{\star} = 0.18$, respectively. This suggest that they may potentially harbor giant planets with sizes of 10.5 $R_{\oplus}$ and 9.75 $R_{\oplus}$. However, they are also large enough that they might be low-mass stars or brown dwarfs as well. We did not include the characterization of EPIC 211995398, because of the low signal-to-noise ratio of its observation. If confirmed to be true planets, these exo-Jupiters may become valuable laboratories for atmospheric characterization for future exoplanet missions. The same is true for planetary systems of bright and nearby hosts, which could be studied in more detail by the next generation of large space observatories, most notably the Transiting Exoplanet Survey Satellite \citep{2014SPIE.9143E..20R} and the James Webb Space Telescope (JWST).

As a first order assessment of habitability, we considered the bulk composition of these planets and whether or not they reside in the habitable zone of their host stars \citep{1979Icar...37..351H,1993Icar..101..108K,2016PhR...663....1S}. \citet{2015ApJ...801...41R} indicate that planets larger than 1.6 $R_{\oplus}$ have densities too low to be rocky or terrestrial, and therefore are likely gaseous. Because we do not have measured masses for these planets, we can merely speculate from their inferred radii that probably only a small fraction of planets in our sample is rocky, based on \citet{2015ApJ...801...41R}.
To address the question of whether the planets are in the habitable zone of their stars, we used the optimistic habitable zone boundaries for M dwarfs presented in \citet{2013ApJ...765..131K}.
They determined that the habitable zone for planets around typical M dwarfs is the circumstellar region between 0.09 and 0.24 AU from the star. Or equivalently, for planets to be considered within the habitable zone of M dwarfs, they must have equilibrium temperatures between 283 K (inner edge of habitable zone) and 171 K (outer edge), assuming an Earth-like, Bond albedo of 0.3. The planets in our sample, however, have very short period orbits ($\overline{P} = 5.9$ days) and live well inside the inner edge of the habitable zone. The proximity of these planets to their host stars render them too hot to be considered habitable, at least from the habitability metrics defined here. We conclude that there are no habitable planets in our sample, since there are none that are both rocky and on the habitable zone (see Figure 6). 

\section{False Positives}
We assess the planetary nature of the candidate planet sample using \textit{vespa}, a statistical validation framework developed by \citet{2012ApJ...761....6M,2015ascl.soft03011M}. This important tool allows the computation of the false positive probability (FPP) of planet candidates by taking their transit and stellar parameters as input. This provides a way to statistically confirm planets for which mass measurements from radial velocity are expensive or not feasible.  

We supported our analysis of each planet candidate in our sample with observations from archival adaptive optics (AO) and speckle images in the Exoplanet Follow-up Observing Program for K2 website (ExoFOP-K2) \footnote[1]{https://exofop.ipac.caltech.edu/k2/}. For EPIC 202071401, the available AO/Speckle images from both Palomar and Keck II show a nearby companion at a separation of $< \sim 3\arcsec$, and even with the smallest aperture, we cannot rule out the possibility of a binary scenario. Moreover, we calculated a high FPP of $7.85\times10^{-1}$ for this object, so we classify it as a planet candidate. Similarly, images from the United Kingdom Infrared Telescope (UKIRT) in ExoFOP-K2 reveal a nearby, bound companion around EPIC 211305568. This early M dwarf has two candidates around it, one of which is classified as a planet candidate in \citet{2017AJ....154..207D} due to their high FPP of $50\%-100\%$ from $vespa$, and due to the nearby companion of the star. It is more probable for systems with two or more candidates to host true planets rather than multiple false positive signals \citep{2012ApJ...750..112L}, so we can apply a boost to the probability of a planet scenario by reducing the FPP of an individual candidate by a factor of 25 for systems with two candidates, or 50 for systems of 3 or more. But even after applying this multiplicity boost to EPIC 211305568, the FPP of $1.74\times10^{-2}$ is too high to statistically confirm it. 
We also reject EPIC 211817229 because the K2 lightcurve shows secondary eclipses, which is suggestive of a binary, although no companion is observed in the ExoFOP-K2 images from UKIRT. The FPP for this candidate is also high, with FPP$=4.02\times10^{-1}$.

EPIC 211509553 is reported in \citet{2016MNRAS.461.3399P} as a candidate with a large transit depth of $R_{p}/R_{}\star = 0.18$. \citet{2017AJ....154..207D} identify it as a cool giant with a period of $P=20.3$ days and radius of $R_{p} = 10.8 \pm 0.6R_{\oplus}$. Although this candidate meets their validation threshold of $< 1\%$ FPP, they cannot statistically confirm it due to the presence of a nearby companion. The available UKIRT and Gemini-8m images show a clear stellar neighbor in the aperture for this star, and we compute an FPP of this candidate of $1.09\times10^{-2}$. We therefore classify it as a planet candidate, in agreement with the literature.

\citet{2016MNRAS.461.3399P} report EPIC 211995398 as a transiting planet candidate with a deep transit of $R_{p}/R_{\star} = 0.19$ and period of $P = 32.5$ days. A recent study found that $~50\%$ of $Kepler$ giant exoplanet candidates are eclipsing binaries \citep{2016A&A...587A..64S}, in contrast to a rate of 18\% from previous study \citep{2013ApJ...766...81F}.
\citet{2013ApJ...766...81F} had found that the $Kepler$ (and K2) false-positive rate depends on planet size, peaking for giants (6-22 $R_{\oplus}$) at 17.7\%. \citet{2016A&A...587A..64S} also demonstrated that many of those false positives turn out to be brown dwarfs \citep{2010ApJ...718.1353I,2011ApJ...730...79J}, with an occurrence rate of $\sim 2\%$. Presently, very few hot Jupiters have been confirmed around M dwarfs \citep{2012AJ....143..111J,2015AJ....149..166H, 2018MNRAS.475.4467B}. So although these large exoplanet candidates (EPIC 211509553 and EPIC 211995398) appear unique, it might be prudent to first rule out that they are false positives with spectroscopic follow-up.

For the rest of the planet candidates for which the K2 photometry and AO/Speckle images were solid enough for $vespa$, we computed their FPP, applying the multiplicity boosts where applicable, and report them in Table 3. As in \citet{2018AJ....155..136M}, we accept as statistically confirmed planets only those candidates with FPP $< 0.001$, and classify anything above that cutoff threshold as a planet candidate. Out of 33 initial candidates, we classify 4 candidates as statistically validated planets and 29 as planet candidates.

\begin{figure}[!ht]
\centering
\includegraphics[width=0.45\textwidth]{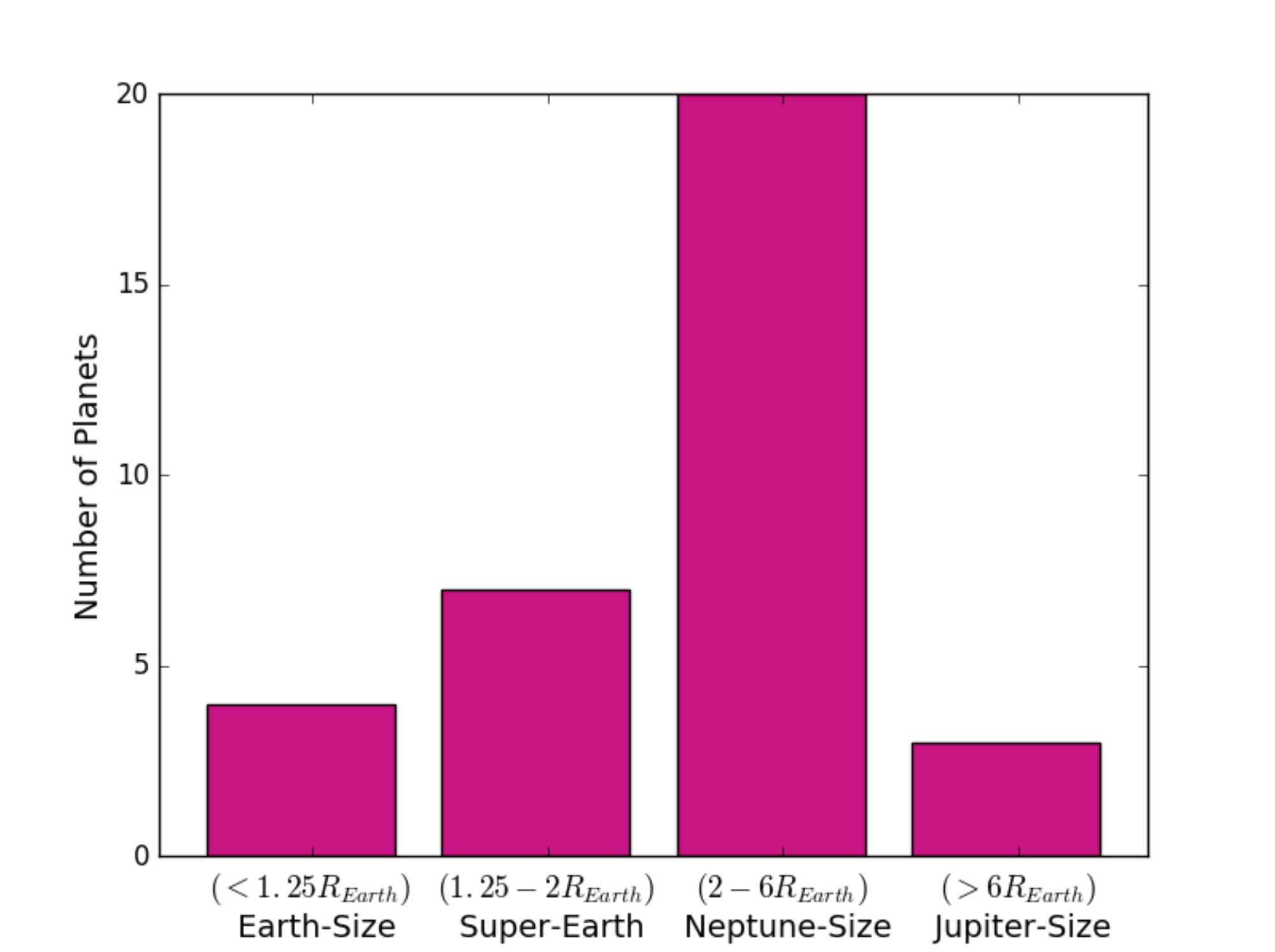} \caption{Planet-size distribution in 4 main categories: Earth size $(< 1.25 R_{\oplus})$, Super-Earth size $(1.25-2 R_{\oplus})$, Neptune-size $(2-6 R_{\oplus}$ and Jupiter-size $(6-15 R_{\oplus})$. The most numerous category is Neptune-size, with 19 exoplanets, over half of the total sample size.} 
\end{figure}

\begin{figure}[!ht]
\centering 
\includegraphics[width=0.45\textwidth]{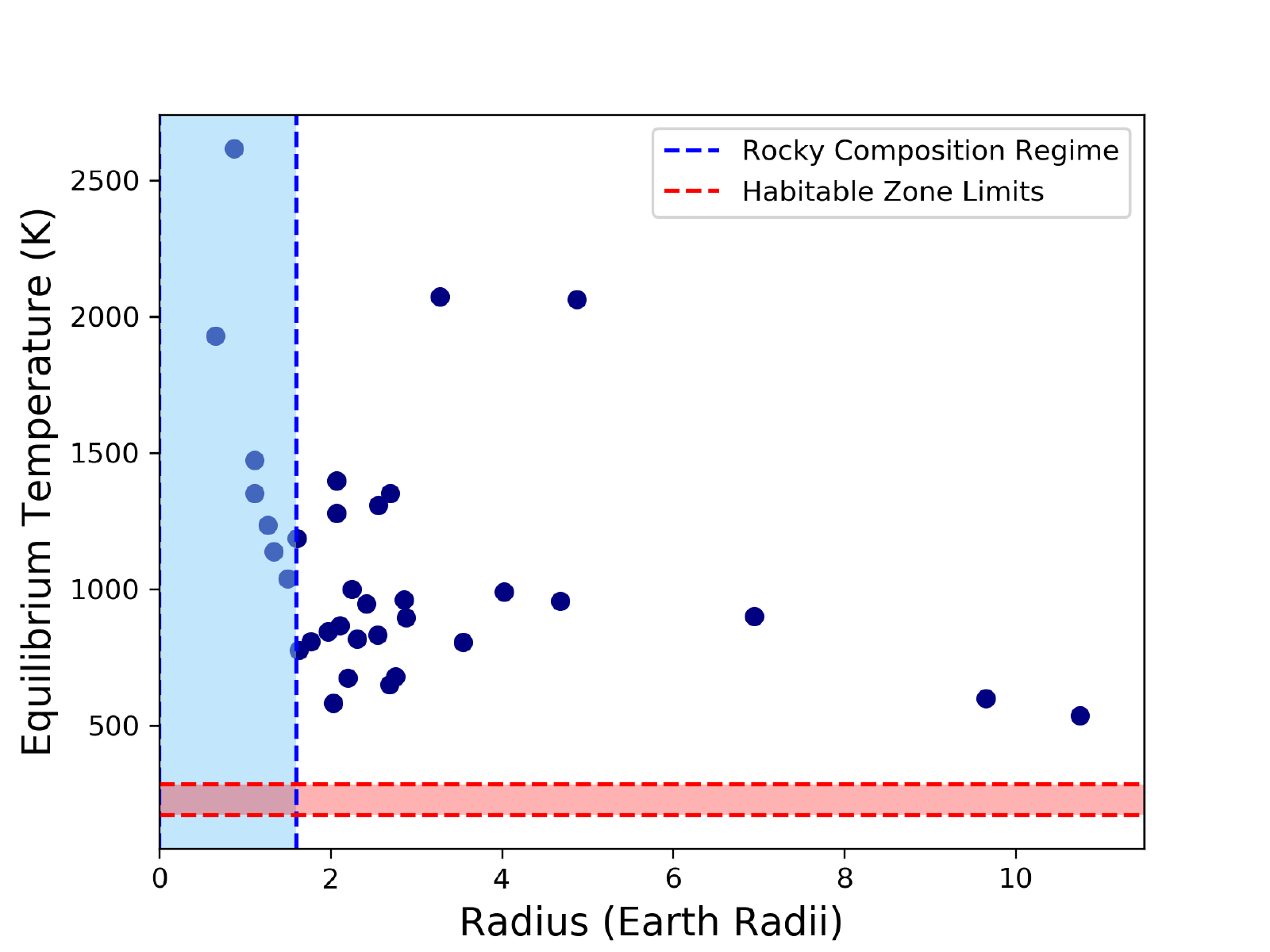} \caption{Distribution of sample of planets by equilibrium temperature versus radius. The blue band defines the region within which planets have rocky compositions, based on \citet{2015ApJ...801...41R}. The red band represents the range of temperatures where planets could be considered temperate for life based on \citet{2013ApJ...765..131K}. Under our definitions of habitability, potentially habitable planets would lie in the overlap of these two bands, the region in which they are both rocky and 
temperate.}
\end{figure}

\section{Conclusions}

In this study, we employ NIR spectra to derive the physical properties of a subset of M dwarf exoplanets and their host stars, uncovered by K2. We adopted a number of empirical calibrations for low-mass stars that relate the EWs of spectral features in the NIR to the stars' physical properties. We compared our EWs of a line sensitive to both radius and temperature, to those from other publications and find that they are in good agreement. Our original sample of K2 stars was contaminated by red giants or dwarfs of hotter classifications, and we discarded those from the characterization presented here.
Additionally, we characterized the associated exoplanet candidates of the stellar sample using the inferred updated properties of their hosts. We specifically estimated the candidate planets' radius and temperature. Our planet sample is largely comprised of small planets, with 11 exoplanet candidates with $R_{p} < 2 R_{\oplus}$, and 22 exoplanets (66\%) with $2 R_{\oplus}< R_{p} <6 R_{\oplus}$. We assessed the habitability of these planets and determined that although some of them might be consistent with a rocky bulk composition, they are too highly radiated by their host stars to be in the habitable zone. Nevertheless, because the stars studied here are relatively bright targets $(\overline{K_{s}} = 11.5)$, some of them could be suitable for follow-up characterization with JWST. In particular, we highlight two systems that are good for atmospheric characterization with HST, $Spitzer$ or JWST: EPIC 211509553 (with R= 9.65 $R_{\oplus}$ and $R_{p}/R_{\star}= 0.18$) which has been statistically validated in other publications as a cool giant; and EPIC 211995398 (R= 10.5 $R_{\oplus}$ and $R_{p}/R_{\star}= 0.15$), which remains a candidate at present. 
Of our final sample of 35 M dwarfs, 24 possess published characterization \citet{2017ApJ...836..167D}, while 11 are new to the literature. These 11 bring the total number of validated exoplanets to 318 from NASA's K2 mission to date.

\acknowledgements

The authors thank Juliette Becker for her assistance with the data reduction using Spextool. We also thank Nick Edwards, Yousef Lawrence, and Jonathan Swift (The Thacher School) for assisting with the data collection at Palomar Observatory. A.V.'s contribution to this study was performed in part under contract with the California Institute of Technology (Caltech)/Jet Propulsion Laboratory (JPL) funded by NASA through the Sagan Fellowship Program executed by the NASA Exoplanet Science Institute. 
Work by B.T.M. was performed in part under contract with the Jet Propulsion Laboratory (JPL) funded by NASA through the Sagan Fellowship Program executed by the NASA Exoplanet Science Institute.

This work has made use of data from the European Space Agency (ESA) mission {\it Gaia} (\url{www.cosmos.esa.int/gaia}), processed by the {\it Gaia}
Data Processing and Analysis Consortium (DPAC,
\url{www.cosmos.esa.int/web/gaia/dpac/consortium}). Funding for the DPAC
has been provided by national institutions, in particular the institutions
participating in the {\it Gaia} Multilateral Agreement.

\newpage

\begin{table*}
\begin{threeparttable}
\caption{Stellar Parameters for K2 Cool Dwarfs}
\centering
\begin{tabular}{ccccccccc}

\hline\hline
EPIC & T$_{\rm{eff}}$ [K] & $\sigma_{T_{\rm{eff}}}$ & Radius [$R_{\odot}$] & $\sigma_{R}$ & Luminosity (log($L_{\star}/L_{\odot}$) & $\sigma_{\rm{log}(L_{\star}/L_{\odot})}$ & [Fe/H] (dex) & $\sigma_{\rm{[Fe/H]}}$ \\
\hline
210564155 & 2870 & 315 & 0.43 & 0.10 & -2.46 & 0.19 & -0.37 & 0.16 \\
211817229 & 2903 & 247 & 0.43 & 0.58 & -2.67 & 0.15 & -0.14 & 0.1 \\
210659688 & 3050 & 122 & 0.41 & 0.04 & -2.3 & 0.1 & 0.13 & 0.1\\
212092746 & 3100 & 84 & 0.14 & 0.05 & -2.03 & 0.14 & 0.27 & 0.15 \\
211839798 & 3258 & 200 & 0.09 & 0.13 & -2.2 & 0.11 & -0.25 & 0.07 \\
210931967 & 3276 & 90 & 0.37 & 0.03 & -1.83 & 0.08 & -0.35 & 0.08 \\
211077024 & 3279 & 68 & 0.29 & 0.02 & -1.9 & 0.08 & 0.4 & 0.05 \\
201205469 & 3299 & 54 & 0.58 & 0.03 & -1.6 & 0.08 & 0.73 & 0.07\\
211916756 & 3315 & 217 & 0.42 & 0.08 & -2.06 & 0.22 & 0.26 & 0.13 \\
212002525 & 3380 & 190 & 0.22 & 0.02 & -2.46 & 0.17 & 0.83 & 0.1 \\
211901114 & 3389 & 114 & 0.4 & 0.02 & -1.9 & 0.1 & 0.37 & 0.12 \\
211428897 & 3393 & 70 & 0.38 & 0.02 & -1.93 & 0.05 & -0.11 & 0.04\\
212154564 & 3429 & 92 & 0.33 & 0.02 & -1.71 & 0.1 & 0.11 & 0.1 \\
210750726 & 3526 & 53 & 0.5 & 0.02 & -1.38 & 0.07 & 0.07 & 0.07 \\
210838726 & 3574 & 46 & 0.52 & 0.01 & -1.48 & 0.07 & 0.03 & 0.06 \\
211305568 & 3618 & 108 & 0.53 & 0.05 & -1.41 & 0.1 & -0.004 & 0.07 \\
210495066 & 3621 & 53 & 0.5 & 0.01 & -1.27 & 0.09 & 0.01 & 0.07 \\
211843564 & 3681 & 86 & 0.54 & 0.01 & -1.45 & 0.16 & 0.48 & 0.13 \\
211969807 & 3720 & 120 & 0.49 & 0.01 & -1.3 & 0.14 & 0.5 & 0.12 \\
201617985 & 3726 & 129 & 0.52 & 0.02 & -1.35 & 0.16 & 0.17 & 0.13 \\
211831378 & 3737 & 68 & 0.54 & 0.01 & -1.3 & 0.13 & 0.26 & 0.12 \\
211924657 & 3766 & 195 & 0.68 & 0.16 & -1.91 & 0.16 & 0.1 & 0.1 \\
211357309 & 3778 & 67 & 0.47 & 0.02 & -1.47 & 0.03 & -0.03 & 0.04 \\
211509553 & 3786 & 107 & 0.49 & 0.02 & -1.21 & 0.12 & -0.18 & 0.13 \\
212006344 & 3837 & 20 & 0.58 & 0.004 & -1.15 & 0.04 & 0.62 & 0.03 \\
211331236 & 3847 & 101 & 0.51 & 0.03 & -1.37 & 0.09 & -0.06 & 0.07 \\
211799258 & 3857 & 212 & 0.38 & 0.02 & -1.72 & 0.11 & 0.26 & 0.11 \\
201833600 & 3911 & 129 & 0.57 & 0.04 & -0.99 & 0.27 & 0.32 & 0.2 \\
211822797 & 4004 & 74 & 0.56 & 0.01 & -1.16 & 0.1 & 0.51 & 0.07 \\
210508766 & 4058 & 314 & 0.72 & 0.09 & -1.69 & 0.67 & 0.61 & 0.33\\
211336288 & 4076 & 77 & 0.57 & 0.01 & -1.4 & 0.12 & -0.39 & 0.08 \\
211762841 & 4105 & 105 & 0.83 & 0.07 & -1.28 & 0.17 & 0.19 & 0.1 \\
201635569 & 4174 & 135 & 0.59 & 0.01 & -0.67 & 0.19 & -0.33 & 0.2 \\
202071401 & 4177 & 87 & 0.71 & 0.04 & -0.73 & 0.09 & -0.49 & 0.07 \\
210968143 & 4187 & 66 & 0.66 & 0.01 & -1.02 & 0.08 & 0.16 & 0.05 \\
  \\
\hline

\end{tabular}
\begin{tablenotes}
\small
\item \textbf{Notes.}\\ The following stars were not included in the characterization either because 1) their observations are too low signal-to-noise to provide reliable estimates of the stellar parameters, or 2) their stellar properties are outside of the bounds of the empirical relationships used here. These are: EPIC 210696763, EPIC
210769880, EPIC 211995398, EPIC 211432922,
EPIC 211694226, EPIC 211826814, EPIC 211970234, EPIC 201155177, EPIC 210524811,
EPIC 210512752, EPIC 210625740, EPIC 212069861, EPIC 212009150, EPIC 211946007, EPIC 212152341, EPIC 211991987, and EPIC 201247497.
\end{tablenotes}
\label{tab:pointings}
\bigskip
\end{threeparttable}

\end{table*}

\begin{table*}
    \caption{Equivalent Width and Masses for K2 cool dwarfs}
    \centering
    \begin{tabular}{ccccccc}
    \hline\hline
    EPIC & Mg ($1.50 \mu m$ ) & Mg ($1.57 \mu m$ ) & Mg ($1.71 \mu m$ ) & Al-a ($1.67 \mu m$ ) & Al-b ($1.67 \mu m$ )  & Mass ($M_{\odot}$) \\
    \hline

210564155 & 0.32 & 2.37 & 1.97 & 1.11 & 1.02 & 0.45 \\
211817229 & 0.74 & 0.70 & 0.97 & 0.31 & 0.75 & 0.45 \\
210659688 & 1.14 & 2.25 & 1.60 & 0.97 & 1.27 & 0.42 \\
212092746 & 2.26 & 0.53 & 1.44 & 1.02 & 2.37 & 0.18 \\
211839798 & 1.63 & 0.40 & 1.52 & 0.39 & 0.86 & 0.36 \\
210931967 & 2.25 & 2.00 & 2.29 & 0.94 & 1.70 & 0.41 \\
211077024 & 2.53 & 1.50 & 1.71 & 1.13 & 1.82 & 0.28 \\
201205469 & 3.99 & 3.76 & 3.30 & 1.13 & 2.79 & 0.61 \\
211916756 & 2.42 & 2.23 & 1.29 & 0.99 & 1.55 & 0.44 \\
212002525 & 1.74 & 1.06 & 0.80 & 1.04 & 1.00 & 0.44 \\
211901114 & 2.29 & 2.46 & 2.44 & 1.30 & 1.50 & 0.26 \\
211428897 & 2.44 & 2.20 & 1.68 & 1.05 & 1.43 & 0.36 \\
212154564 & 3.01 & 1.77 & 2.34 & 1.30 & 1.84 & 0.35 \\
210750726 & 4.32 & 3.26 & 2.61 & 1.22 & 2.19 & 0.51 \\
210838726 & 4.47 & 3.92 & 3.29 & 1.77 & 2.63 & 0.54 \\
211305568 & 4.13 & 2.98 & 2.48 & 0.95 & 1.76 & 0.58 \\
210495066 & 4.82 & 3.17 & 2.85 & 2.04 & 2.95 & 0.52 \\
211843564 & 4.77 & 4.07 & 3.40 & 2.01 & 2.63 & 0.44 \\
211969807 & 4.63 & 3.36 & 2.62 & 1.75 & 2.27 & 0.51 \\
201617985 & 4.89 & 3.76 & 3.27 & 1.52 & 2.18 & 0.55 \\
211831378 & 5.92 & 4.15 & 3.80 & 1.59 & 2.65 & 0.57 \\
211924657 & 3.47 & 2.82 & 0.93 & 0.69 & 1.19 & 0.53 \\
211357309 & 4.02 & 2.42 & 2.80 & 0.84 & 1.40 & 0.55 \\
211509553 & 5.24 & 3.40 & 3.12 & 1.39 & 2.13 & 0.52 \\
212006344 & 7.01 & 5.13 & 4.08 & 1.85 & 3.03 & 0.62 \\
211331236 & 5.14 & 3.42 & 3.50 & 1.22 & 1.83 & 0.63 \\
211799258 & 2.99 & 2.11 & 2.24 & 1.53 & 1.16 & 0.51 \\
201833600 & 7.67 & 4.38 & 4.09 & 1.54 & 2.86 & 0.60 \\
211822797 & 6.48 & 4.39 & 3.82 & 1.60 & 2.22 & 0.63 \\
210508766 & 7.68 & 6.94 & 4.91 & 2.19 & 3.10 & 0.65 \\
211336288 & 6.30 & 4.63 & 4.16 & 1.75 & 2.11 & 0.64 \\
211762841 & 7.34 &5.37 & 4.35 & 1.07 & 2.04 & 0.60 \\
201635569 & 7.25 & 4.98 & 2.97 & 1.98 & 2.31 & 0.61 \\
202071401 & 7.64 & 5.29 & 3.56 & 1.26 & 2.14 & 0.70 \\
210968143 & 7.93 & 5.89 & 4.26 & 1.59 & 2.35 & 0.73 \\

  
    \hline
    \end{tabular}
    \label{tab:pointings}
\bigskip
\end{table*}

\begin{table*}
    \caption{Planet Parameters for K2 Cool Dwarfs}
    \centering
    \begin{tabular}{ccccccccccc}
    \hline\hline
    EPIC & $R_{p}/R_{\star}$ & $R_{p}$ ($R_{\oplus}$)  & $\sigma_{R_{p}}$ & P (days) & $a$ (AU) & $T_{\rm{eq}}$ (K) & VESPA FPP & Multiplicity  & Adopted FPP & Designation \\
     & & & & & & & & Boost & & \\ 
    \hline
210495066 & 0.027 & 1.5 & 0.01 & 3.74 & 0.035 & 1038 & $6.05\times 10^{-2}$ & N & $6.05\times 10^{-2}$ & Candidate \\
210508766 & 0.028 & 2.25 & 0.02 & 2.74 & 0.033 & 1000 & $8.44\times 10^{-3}$ & Y & $3.37\times 10^{-4}$ & Planet \\
210508766 & 0.034 & 2.69 & 0.02 & 9.99 & 0.078 & 650 & $4.90\times 10^{-1}$ &Y&$1.96\times 10^{-2}$& Candidate\\
210659688 & 0.063 & 2.86 & 0.01 & 2.35 & 0.026 & 961 & $1.02\times 10^{-1}$ & N & $1.02\times 10^{-1}$ & Candidate \\
210750726 & 0.044 & 2.42 & 0.01 & 4.61 & 0.04 & 946 &$1.00\times10^{-2}$&N&$1.00\times10^{-2}$&Candidate  \\
210838726 & 0.019 & 1.11 & 0.01 & 1.09 & 0.01 & 1472 &$1.85\times10^{-1}$& N&$1.85\times10^{-1}$ & Candidate \\
210968143 & 0.038 & 2.76 & 0.01 & 13.73 & 0.10 & 678 & $1.21\times10^{-3}$& Y& $4.84\times10^{-5}$& Planet \\
210968143 & 0.018 & 1.34 & 0.02 & 2.90 & 0.035 & 1139 &$2.31\times10^{-1}$&Y& $9.24\times 10^{-3}$& Candidate  \\
210931967 & 0.081 & 3.28 & 0.07 & 0.34 &  0.007 & 2072 &$9.95\times10^{-1}$& N&$9.95\times10^{-1}$ & Candidate \\
211077024 & 0.035 & 1.11 & 0.01 & 1.41 & 0.016 & 1351 &$1.12\times10^{-1}$&N&$1.12\times10^{-1}$& Candidate  \\
202071401 & 0.020 & 1.61 & 0.02 & 3.23 & 0.038 & 1186 &$7.85\times10^{-1}$&N&$7.85\times10^{-1}$&Candidate \\
211305568 & 0.038 & 2.20 & 0.02 & 11.55 & 0.08 & 675 \\
211305568 & 0.015 & 0.87 & 0.02 & 0.19 & 0.005 & 2618 &$4.35\times10^{-1}$&Y&$1.74\times10^{-2}$&Candidate  \\
211331236 & 0.037 & 2.07 & 0.01 & 1.29 & 0.02 & 1399 &$4.98\times10^{-3}$&Y&$1.99\times10^{-4}$& Planet\\
211331236 & 0.038 & 2.11 & 0.01 & 5.44 & 0.05 & 866 &$7.09\times10^{-2}$ &Y&$2.83\times10^{-3}$&Candidate \\
211509553 & 0.180 & 9.65 & 0.00 & 20.35 & 0.118 & 600 &$1.09\times10^{-2}$& N &$1.09\times10^{-2}$&Candidate  \\
211762841 & 0.029 & 2.70 & 0.03 & 1.56 & 0.022 & 1351 &$4.40\times10^{-1}$&N&$4.40\times10^{-1}$&Candidate  \\
211817229 & 0.061 & 2.88 & 0.04 & 2.17 & 0.025 & 896 &$4.02\times10^{-1}$&N&$4.02\times10^{-1}$& Candidate  \\
211822797 & 0.033 & 2.03 & 0.01 & 21.16 & 0.128 & 581 &$1.79\times10^{-3}$&N&$1.79\times10^{-3}$& Candidate \\
211843564 & 0.082 & 4.88 & 0.01 & 0.452 & 0.008 & 2064 &$4.27\times10^{-1}$&N&$4.27\times10^{-1}$&Candidate\\
211901114 & 0.058 & 2.56 & 0.02 & 1.56 & 0.017 & 1308 &$3.18\times10^{-1}$&N&$3.18\times10^{-1}$&Candidate\\ 
211916756 & 0.077 & 3.55 & 0.01 & 10.13 & 0.042 & 805 &$1.35\times10^{-2}$&N&$1.35\times10^{-2}$ & Candidate \\
201635569 & 0.107 & 6.95 & 0.02 & 8.36 & 0.068 & 900 &$5.22\times10^{-2}$&N&$5.22\times10^{-2}$& Candidate\\
201833600 & 0.031 & 1.97 & 0.01 & 8.75 & 0.066 & 845 &$5.06\times10^{-2}$&N&$5.06\times10^{-2}$& Candidate \\
201617985 & 0.031 & 1.77 & 0.02 & 7.28 & 0.06 & 807 &$9.92\times10^{-1}$&N&$9.92\times10^{-1}$ & Candidate \\
210564155 & 0.034 & 1.63 & 0.01 & 4.86 & 0.037 & 776 &$5.23\times10^{-2}$&N&$5.23\times10^{-2}$&Candidate\\
212006344 & 0.020 & 1.27 & 0.01 & 2.21 & 0.028 & 1236 &$1.89\times10^{-3}$&N&$1.89\times10^{-3}$& Candidate\\
212092746 & 0.043 & 0.66 & 0.00 & 0.56 & 0.007 & 1930 &$1.49\times10^{-1}$&N&$1.49\times10^{-1}$&Candidate \\
211969807 & 0.038 & 2.07 & 0.03 & 1.97 & 0.02 & 1280 &1.00&N&1.00&Candidate\\
211924657 & 0.054 & 4.03 & 0.03 & 2.64 & 0.03 & 990 &$3.27\times10^{-1}$&N&$3.27\times10^{-1}$&Candidate\\
212154564 & 0.071 & 2.55 & 0.00 & 6.41 & 0.05 & 832 &$4.38\times10^{-4}$&N&$4.38\times10^{-4}$& Planet\\
201205469 & 0.074 & 4.68 & 0.02 & 3.47 & 0.037 & 956 &$1.95\times10^{-1}$ &N&$1.95\times10^{-1}$&Candidate \\
211799258 & 0.259 & 10.75 & 0.01 & 19.53 & 0.11 & 535 &$8.22\times10^{-1}$&N&$8.22\times10^{-1}$&Candidate \\

  
    \hline
    \end{tabular}
    \label{tab:pointings}
\end{table*}

\newpage

\begin{figure*}
\centering
\includegraphics[width=0.35\textwidth]{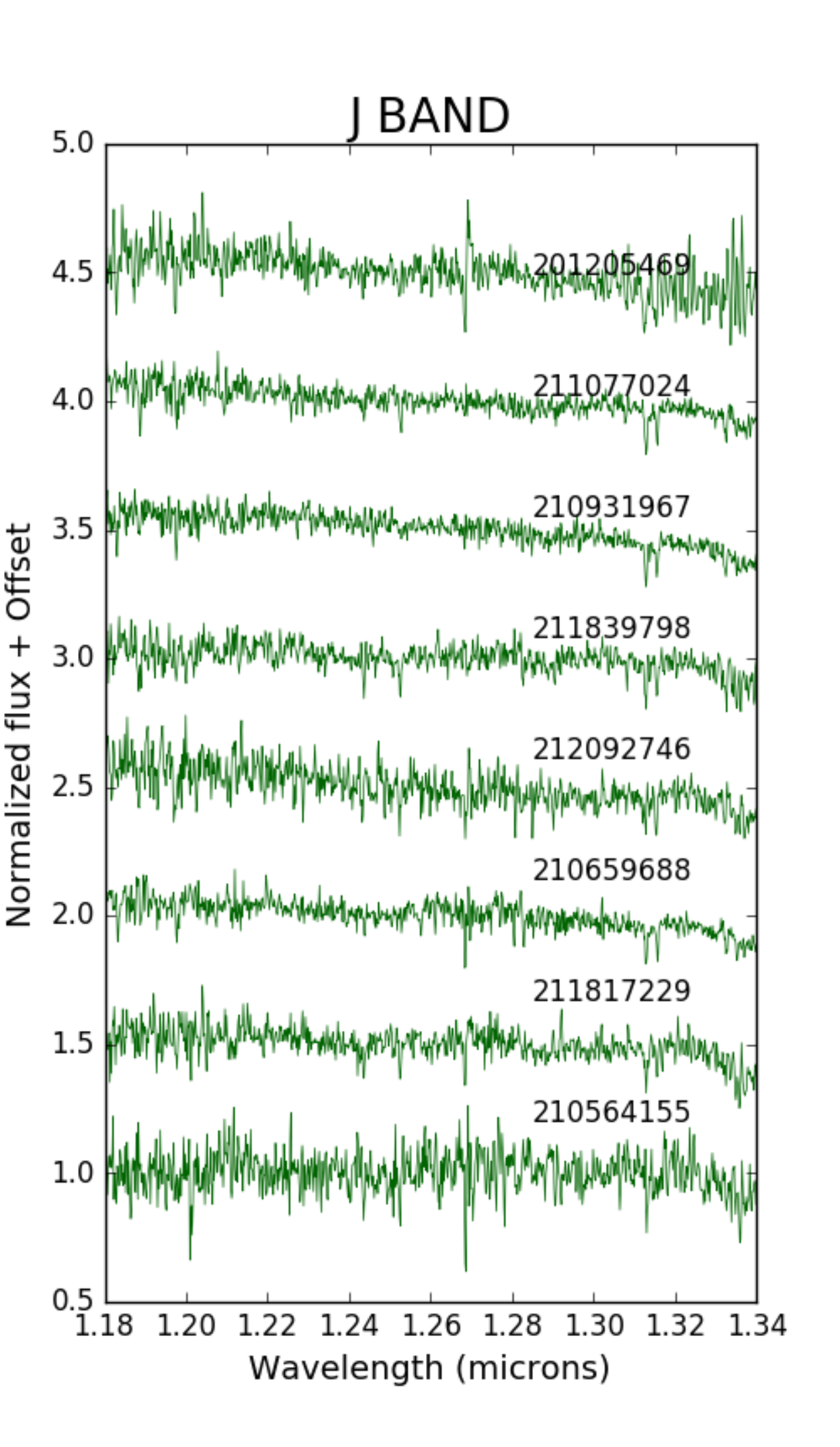}
\includegraphics[width=0.35\textwidth]{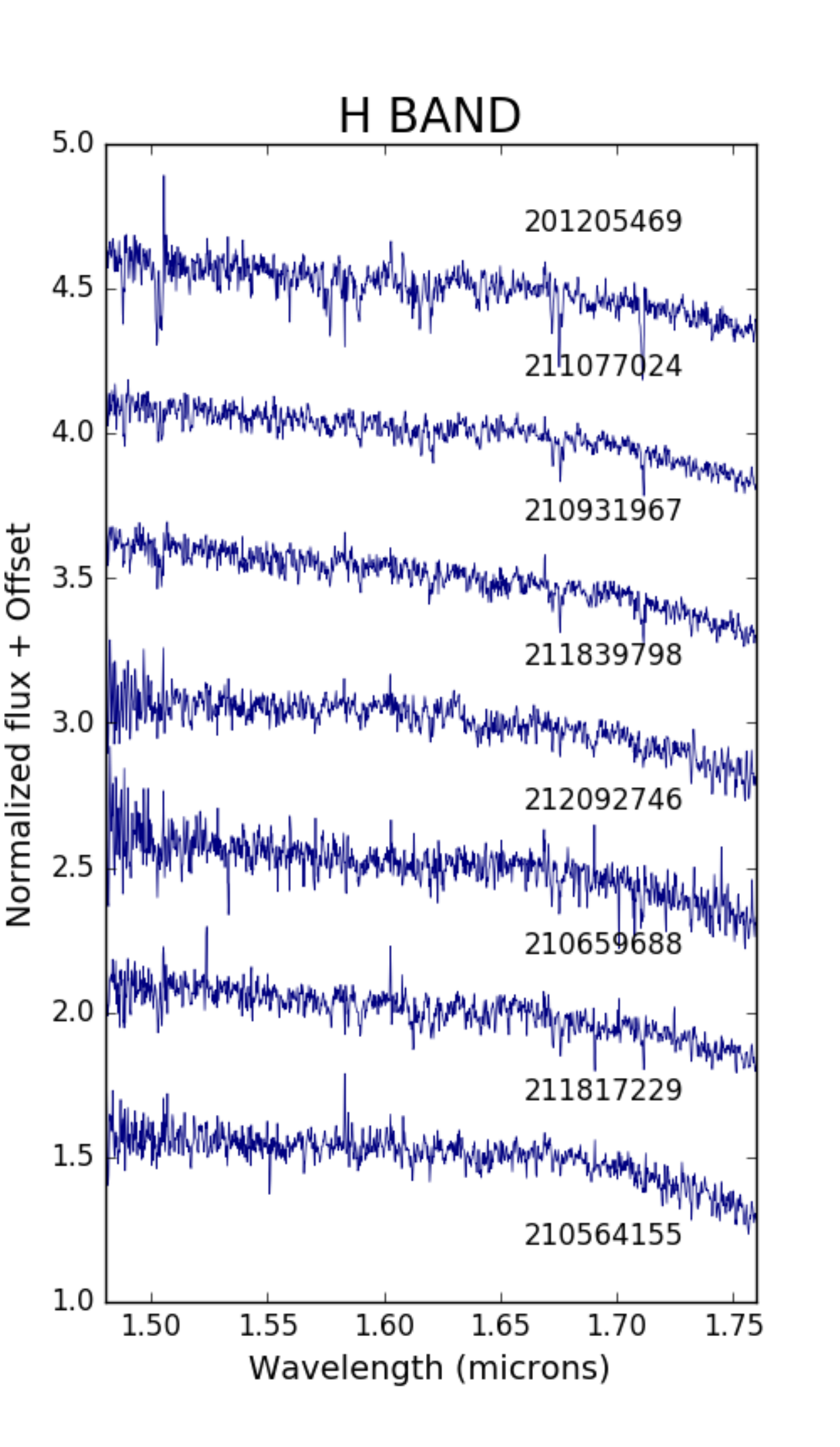} \\
\includegraphics[width=0.35\textwidth]{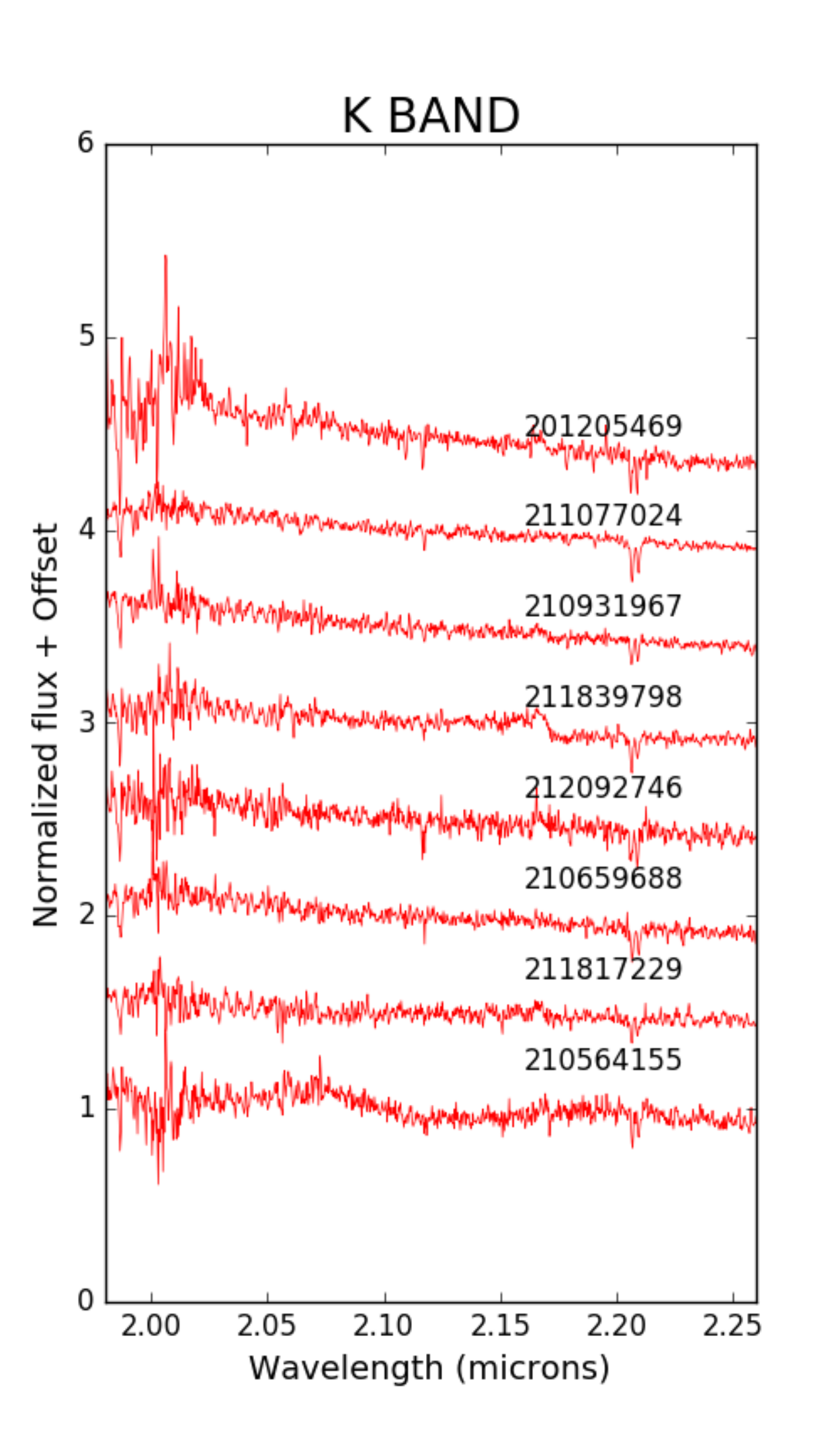}
\includegraphics[width=0.35\textwidth]{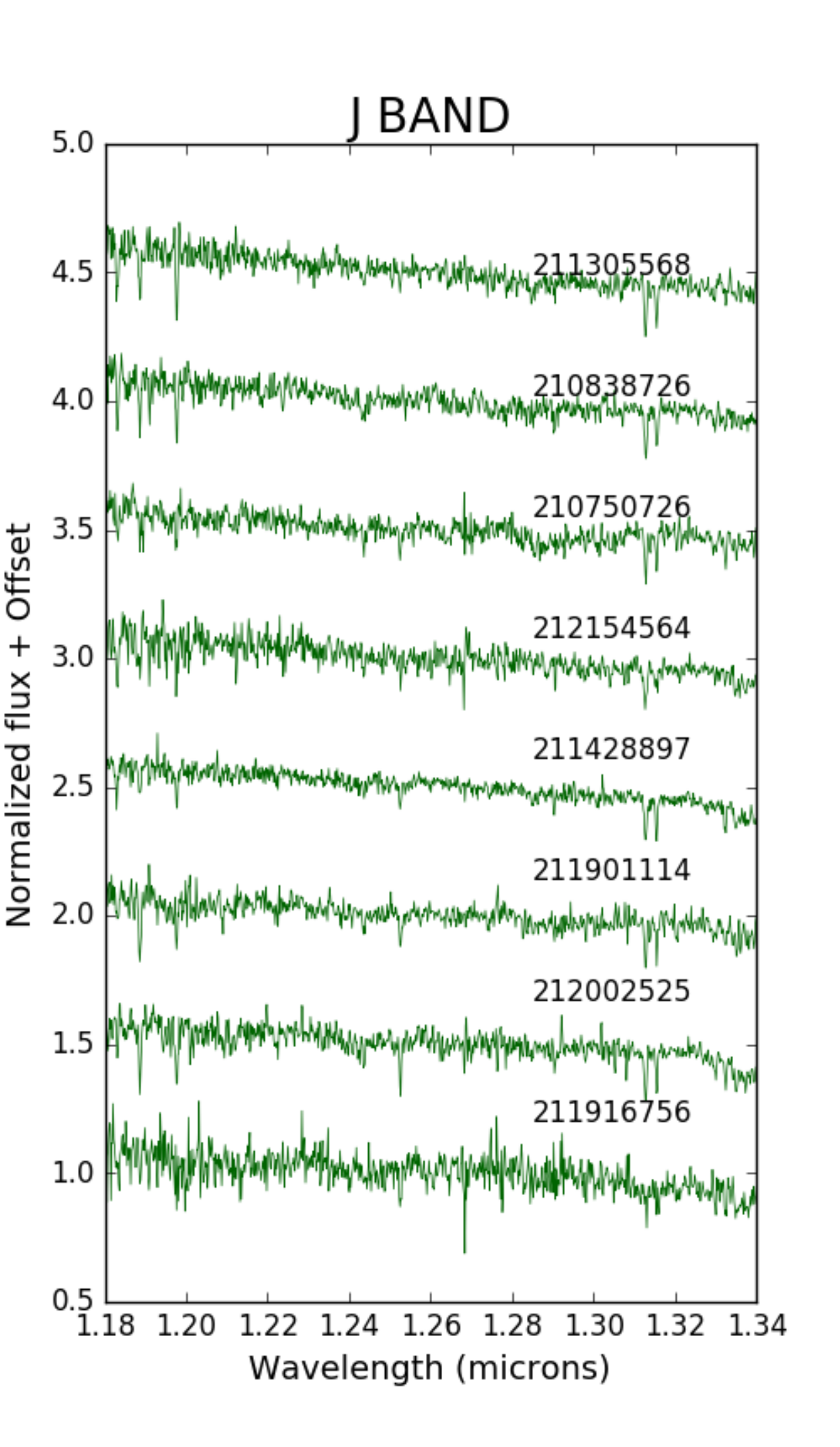}
\caption{Multiple-band spectra of our cool dwarf sample. $J$ band (top left), $H$ band (top right) and $K$ band (bottom left). We are showing the spectra in $H,J,K$ bands in order of decreasing temperature such that the hottest stars are on the top.}
\end{figure*}
\newpage

\begin{figure*}
\centering
\includegraphics[width=0.35\textwidth]{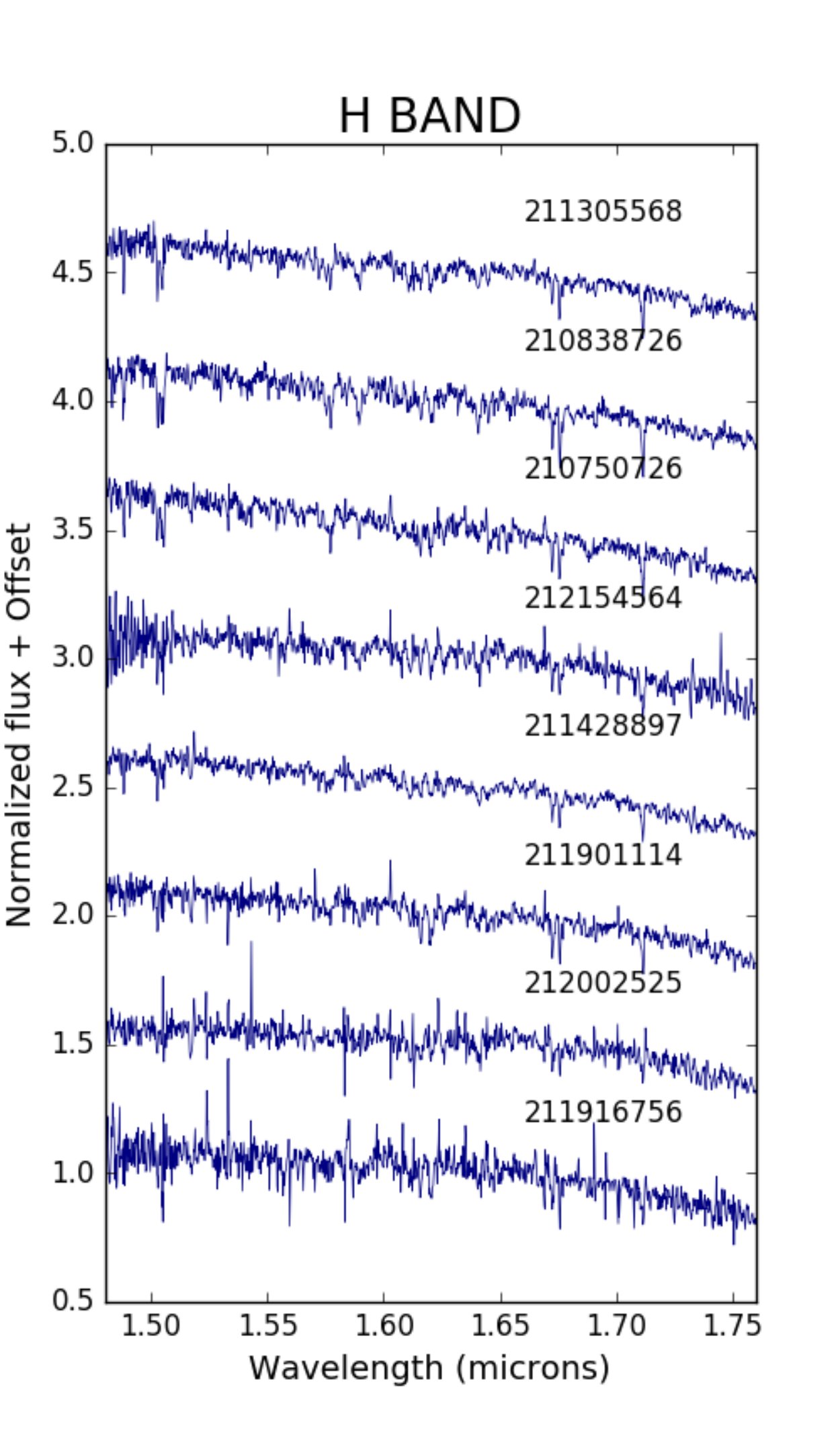}
\includegraphics[width=0.35\textwidth]{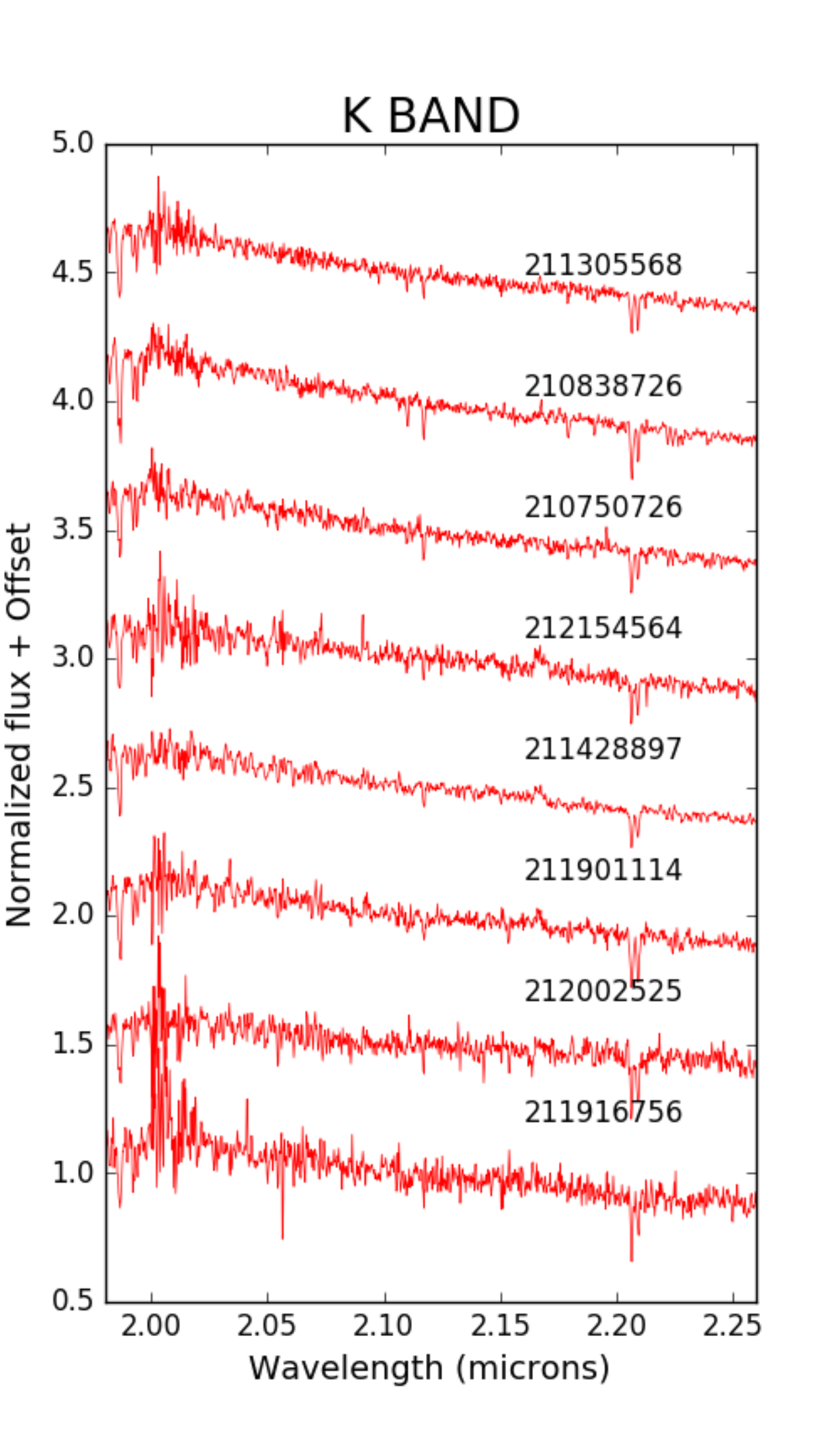} \\
\includegraphics[width=0.35\textwidth]{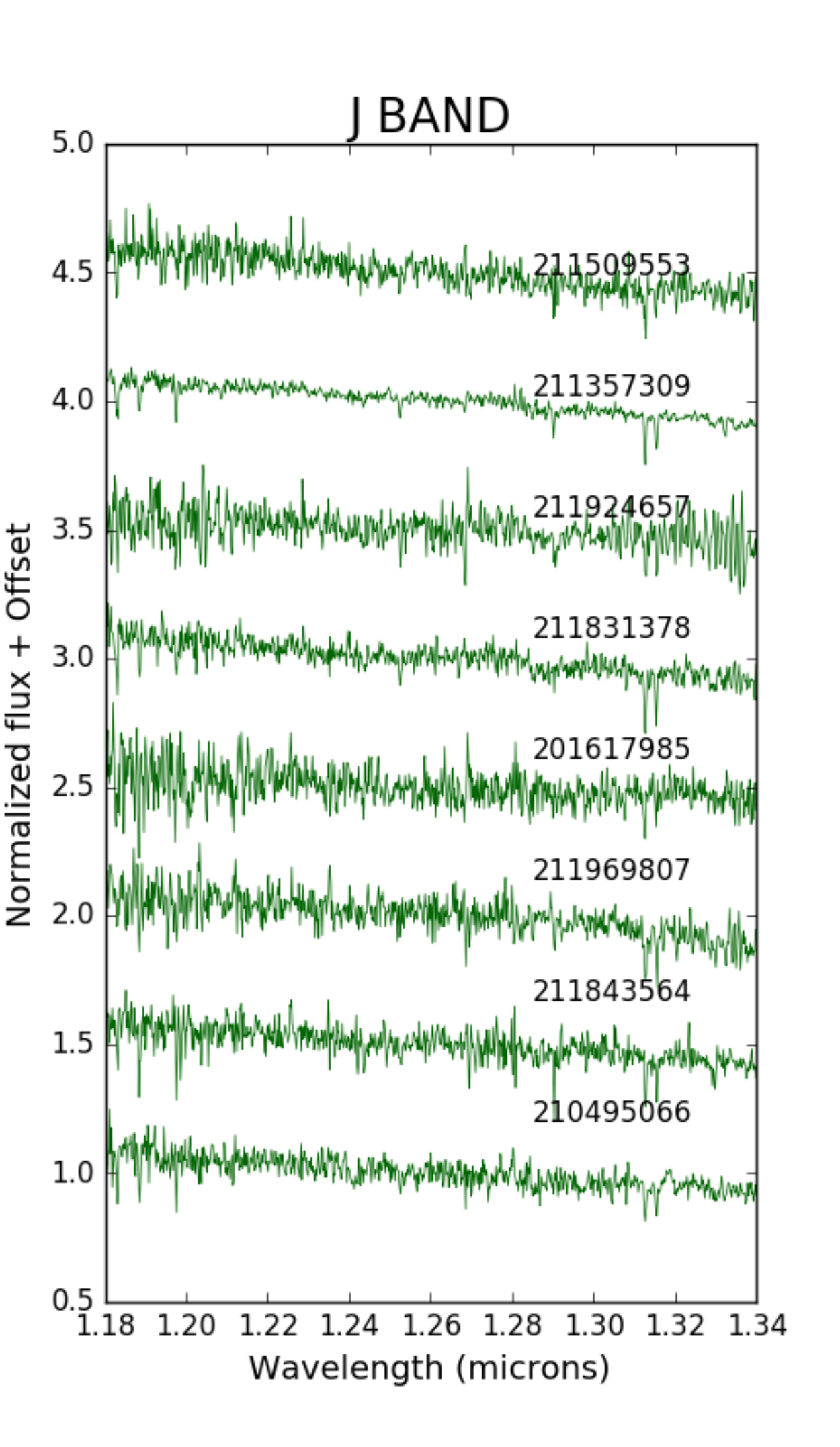}
\includegraphics[width=0.35\textwidth]{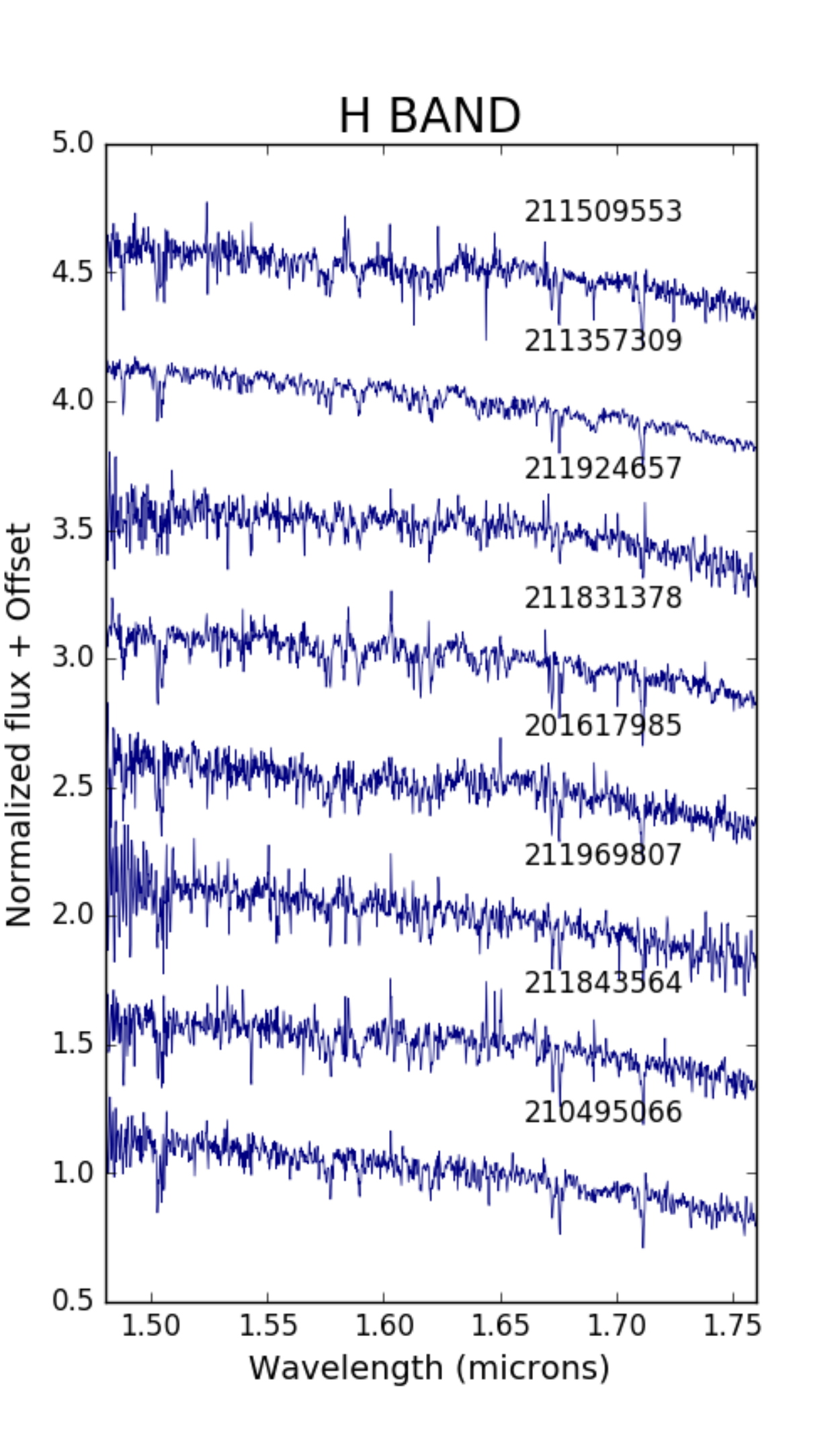}
\caption{Same as Figure 6}
\end{figure*}
\newpage

\begin{figure*}
\centering
\includegraphics[width=0.35\textwidth]{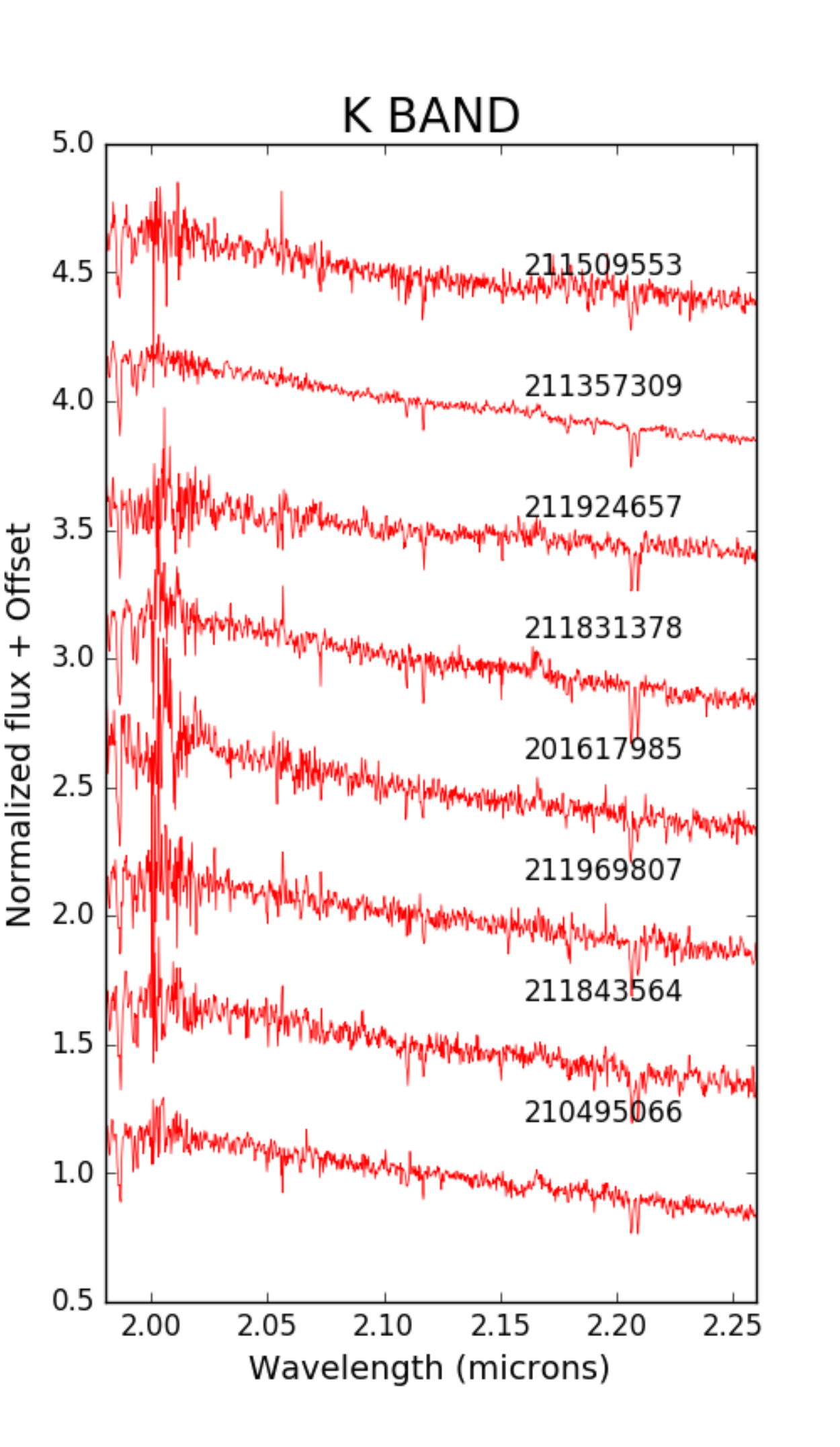}
\includegraphics[width=0.35\textwidth]{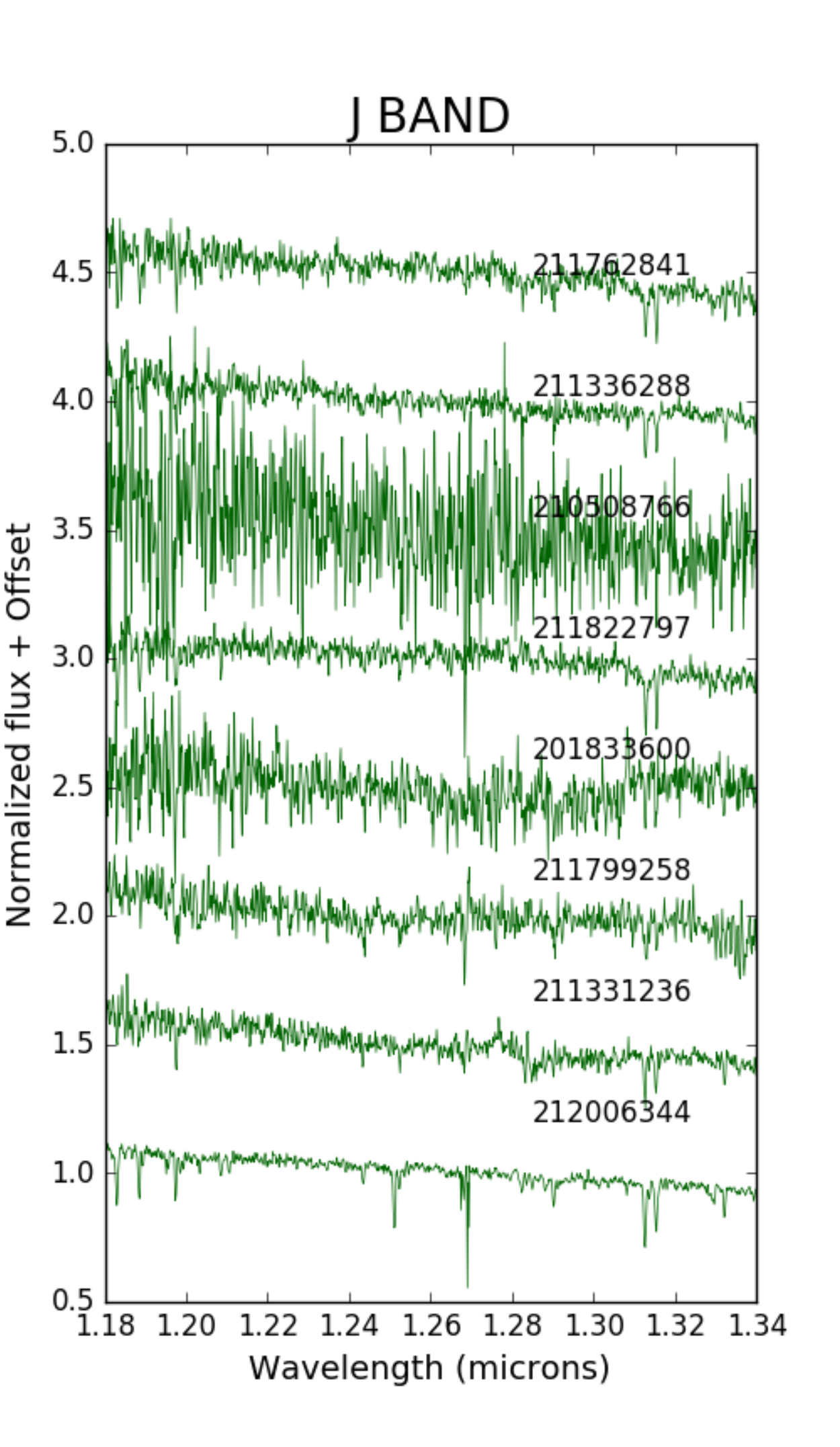}\\
\includegraphics[width=0.35\textwidth]{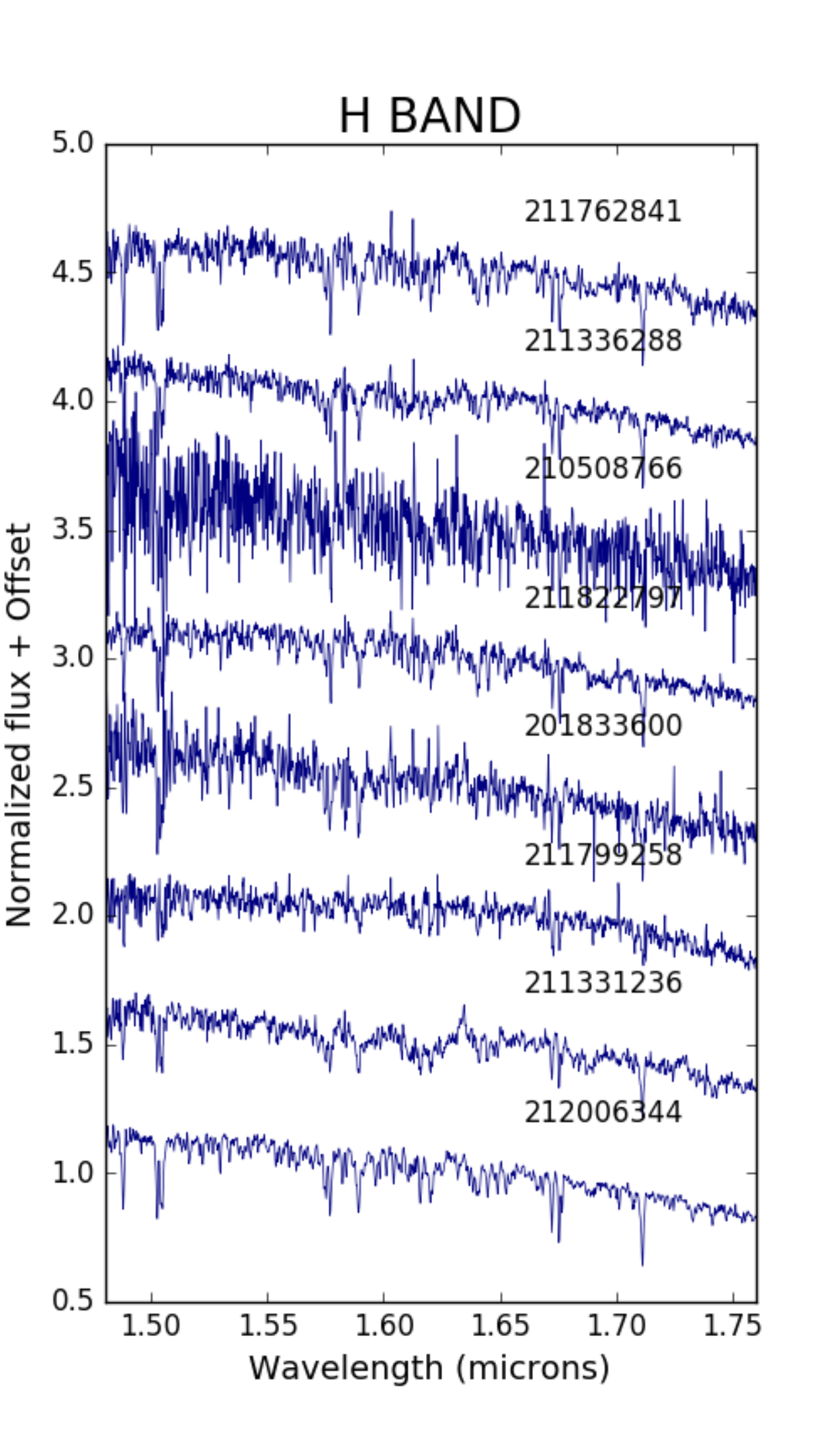} 
\includegraphics[width=0.35\textwidth]{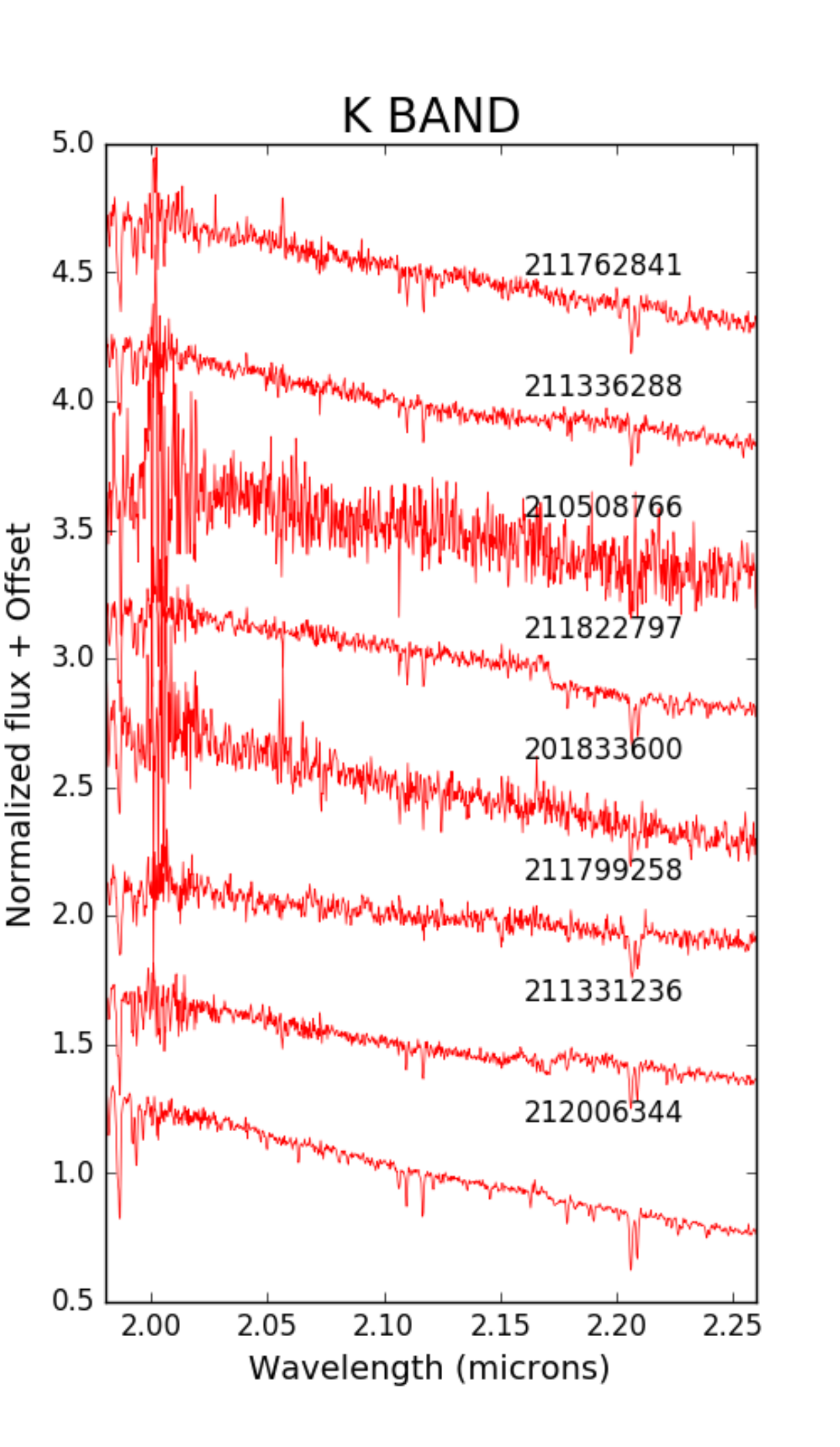}
\caption{Same as Figure 6}
\end{figure*}
\newpage

\begin{figure*}
\centering
\includegraphics[width=0.35\textwidth]{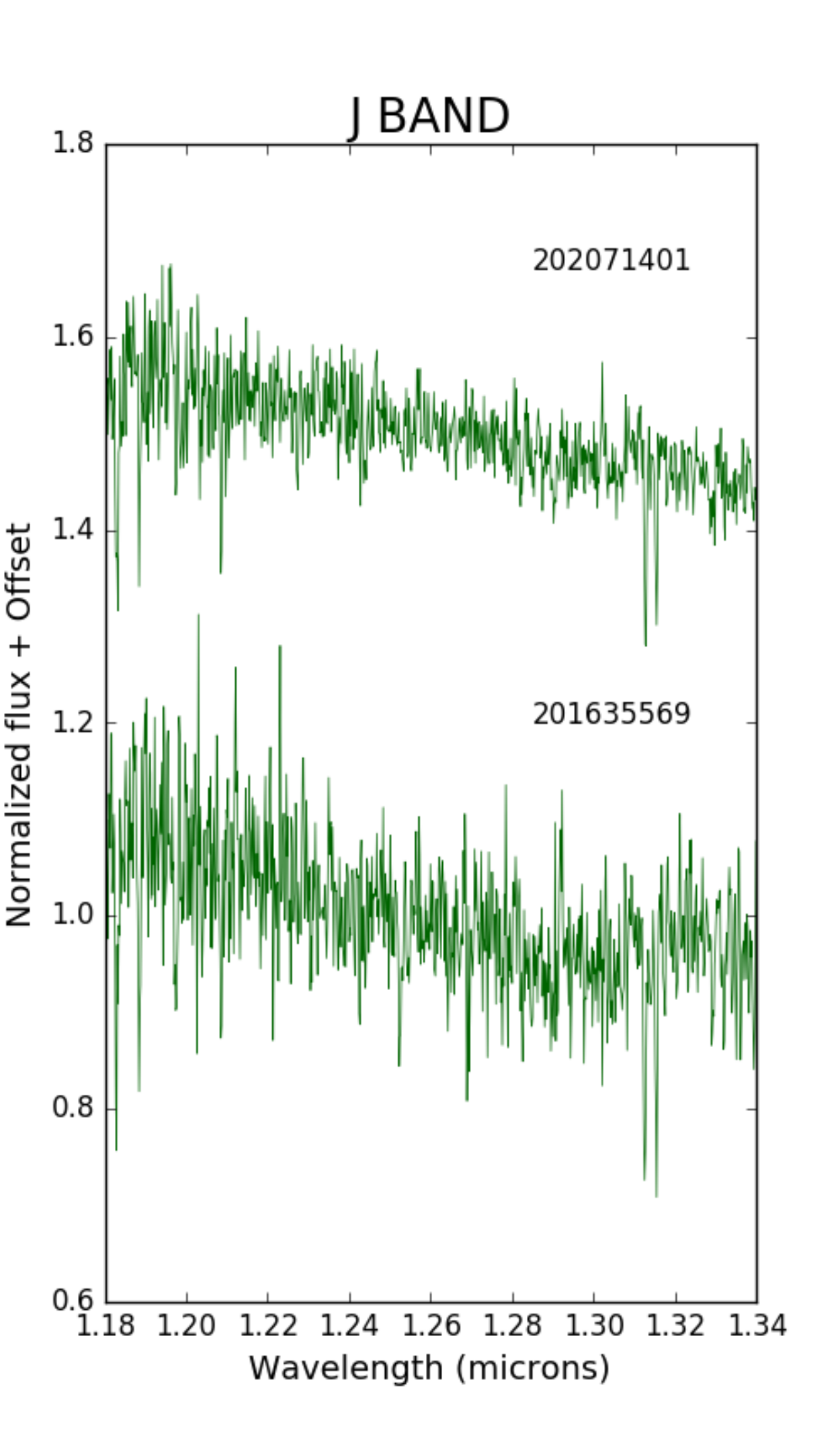}
\includegraphics[width=0.35\textwidth]{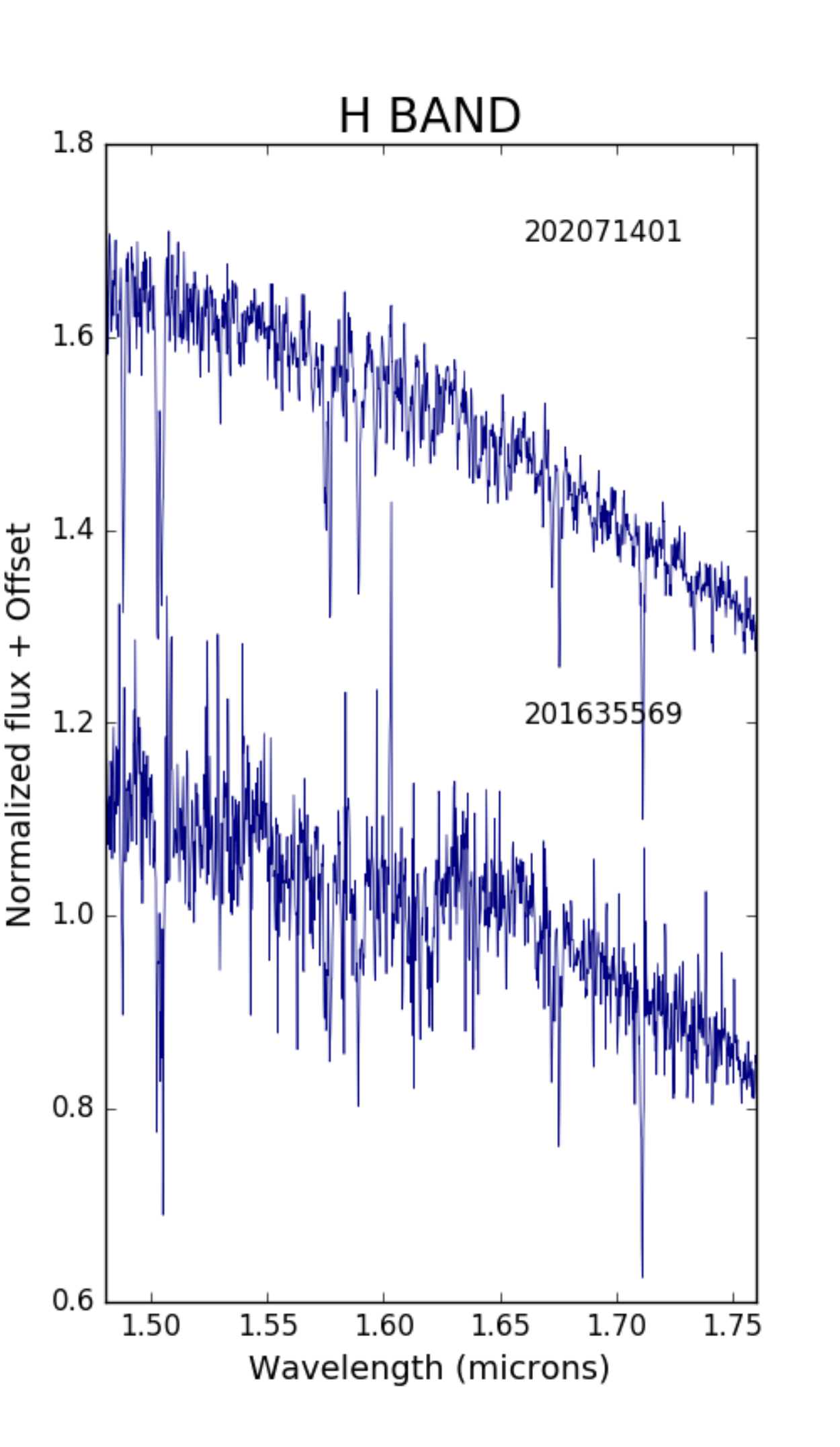} \\
\includegraphics[width=0.35\textwidth]{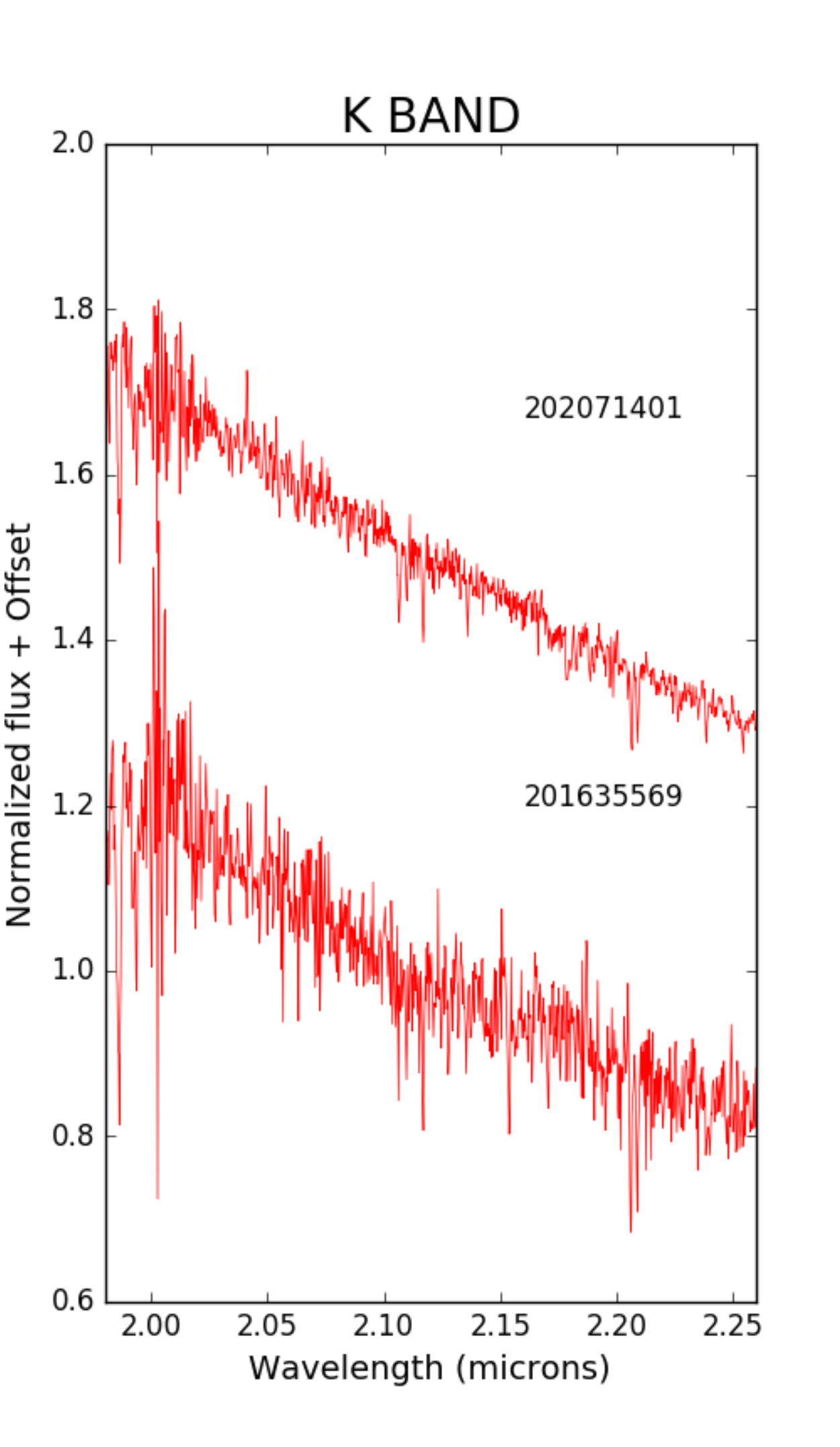}
\caption{Same as Figure 6}
\end{figure*}
\end{document}